\documentclass[aps,prl,reprint,superscriptaddress, longbibliography]{revtex4-1}
\usepackage{graphicx}
\usepackage{graphics}
\usepackage{epsfig}
\usepackage{epstopdf}
\usepackage{placeins}
\usepackage{lipsum}% http://ctan.org/pkg/lipsum

\usepackage{array} %is required
\usepackage{multirow}

\usepackage{xcolor}
\usepackage[colorlinks,linkcolor=blue,citecolor=blue,urlcolor=blue] {hyperref}
\usepackage{amsmath}

\newcommand{\SST}{$\mathrm{Sc_2TbN@C_{80}}$}
\newcommand{\Csix}{$\mathrm{C_{60}}$}
\newcommand{\Ceig}{$\mathrm{C_{80}}$}
\newcommand{\hBN}{$h$-{BN}}
\newcommand{\YSS}{Sc$_2$YN@C$_{80}$}
\newcommand{\SSS}{Sc$_3$N@C$_{80}$}
\newcommand{\SSY}{Sc$_2$YN@C$_{80}$}


\date{\today}

\begin{document}

\title{Temperature induced change in conformation of \SST\ on $h$-BN/Ni(111)
}

\author{Roland~Stania} 
\affiliation{Physik-Institut, Universit\"at Z\"urich, CH-8057 Z\"urich, Switzerland}
\affiliation{Center for Artificial Low Dimensional Electronic Systems,
Institute for Basic Science, KR-37673 Pohang, South Korea}
%------------------------------
\author{Ari Paavo Seitsonen} 
\affiliation{D\'{e}partement de Chimie, \'{E}cole Normale Sup\'{e}rieure, F-75005 Paris, France}
%------------------------------
\author{Hyunjin~Jung} 
\affiliation{Center for Artificial Low Dimensional Electronic Systems,
Institute for Basic Science, KR-37673 Pohang, South Korea}
%------------------------------
\author{David~Kunhardt}
\affiliation{Leibniz Institute of Solid State and Materials Research, D-01069 Dresden, Germany}
%-----------------------------
\author{Bernd~B\"uchner}
\affiliation{Leibniz Institute of Solid State and Materials Research, D-01069 Dresden, Germany}
%-----------------------------
\author{Alexey A.~Popov}
\affiliation{Leibniz Institute of Solid State and Materials Research, D-01069 Dresden, Germany}
%-----------------------------
\author{Matthias~Muntwiler} 
\affiliation{Paul Scherrer Institut, CH-5232 Villigen, Switzerland}
%-----------------------------
\author{Thomas~Greber} 
\affiliation{Physik-Institut, Universit\"at Z\"urich, CH-8057 Z\"urich, Switzerland}
\email{greber@physik.uzh.ch}

%\begin{tocentry}
%   \includegraphics[scale=0.31]{TOCC1s.pdf}
%    \label{fig:TOC}
%\end{tocentry}

\begin{abstract}
The conformation of molecules on surfaces is decisive for their functionality. In the case of the {\mbox{endofullerene}} paramagnet \SST\ the conformation is linked to an electric and a magnetic dipole moment. Therefore a workfunction change of a substrate with such molecules on the surface suggests the system to be magnetoelectric. 
Here one monolayer of \SST\ has been studied on \hBN /Ni(111). The molecules assume a hexagonally close packed lattice aligned with the substrate high symmetry directions. The structure is incommensurate and arranges at a periodicity of $\approx$4.3$\times$4.3 substrate unit cells. At low temperatures a C$_{80}$ ($2\times2$) superstructure is observed. 
Valence band photoemission spectroscopy shows a temperature induced  0.3 eV shift on the C$_{80}$ molecular orbitals to lower binding energies that is parallel to a workfunction increase. From comparison of the molecular orbital angular photoemission intensity distributions it is conjectured that the molecules undergo a change in conformation between 30 and 300 K. 
This phase transition is centred at 125~K as observed with high resolution x-ray photoelectron spectroscopy that shows the core levels of the atomic species on the molecules to shift parallel to the workfunction. 
The temperature dependence can be described with a two level model that accounts for disordering with an excitation energy of 74 meV into a high entropy ensemble.
The experimental findings are backed by density functional theory for the diamagnetic sibling of \SST : \SSY . The calculations rationalize the incommensurate structure, indicate a permanent electric dipole moment of \SSY\, and predict the room temperature N~1s x-ray photoelectron spectrum.  
\end{abstract}
\maketitle

\section{Introduction}
Fullerenes are  archetypal molecular building blocks in nanoscience. Correspondingly carbon shells like C$_{60}$ have been studied in detail on surfaces \cite{rud99,san09,mor101}.  
Like found in three dimensions \cite{hei91}, the freezing of rotational degrees of freedom also cause phase transitions in two-dimensional systems \cite{ben93,gol96,gol02}.
A peculiar phase transition of a monolayer of \Csix\ on a single layer of {\mbox{\hBN\ on Ni(111)}} showed a temperature dependent charge transfer onto \Csix\ that is triggered by a change in conformation of the molecules with respect to the 
substrate \cite{mun05}.
This pointed to non-adiabatic effects that may be important for the engineering of electronic or spintronic contacts at the nanometer scale. 

{\it{Endohedral}} fullerenes are carbon cages that contain atoms \cite{pop13}.
Species with a similar robustness as \Csix\  \cite{ste99} open more opportunities for the exploitation of molecular functionality such as single molecule magnetism \cite{wes12}. 
First surface science experiments focused on the electronic properties of multilayer \Ceig\ endofullerenes \cite{alv02, shi05}, and soon monolayer systems were prepared and observed with scanning probes \cite{lei07,tre09,tia11,kry20}.
At low temperatures the orientation of the endohedral unit is not random \cite{tre09} and related to the magnetisation \cite{wes15,gre19,kry20}.
The orientation of the endohedral units may even be changed by magnetic fields \cite{kos17}.
On the other hand, the incomplete screening of the anisotropic electrostatic potential of the endohedral cluster \cite{sta18} bears a handle for accessing the magnetisation of the endohedral clusters with electric fields.

Here we report on a phase transition of  \SST\ on \hBN/Ni(111) around 125~K. It is related to the change in molecular conformation which is reflected in a change of the angular dependence of the {\mbox{photoemission}} from molecular orbitals and accompanied by a change of the workfunction, as was the case for {\mbox{\Csix\ on \hBN /Ni(111)}}. In the present transition, however, we find no charge transfer onto the molecule. Instead, the orbital energies of \SST\ shift parallel to the workfunction i.e.\ they align with the vacuum level.

%***********************************************************************

\section*{Experimental and theoretical details}

\subsection*{Experimental}

TbSc$_2$N@C$_{80}$ molecules with icosahedral $I_h$ symmetry of the carbon cage were synthetized and purified as described in Ref.~\cite{zha15} and sublimated onto \hBN/Ni(111) \cite{auw99}, with the substrate kept at 470~K. Like in the case of C$_{60}$ \cite{mun05} this substrate does bind the molecules weakly. The photoemission data were recorded at a photon energy of 600~eV or He~I$\alpha$ (21.2 eV) \cite{mun17}. The coverage was determined with a layer by layer growth model, with an electron mean free path of 1 nm and from the intensity ratio of the N~1s core levels of the molecule and {\mbox{\hBN\ }} \cite{sta18}.
The low energy electron diffraction (LEED) patterns were calibrated with the (0,1) spots of the $h$-BN/Ni(111) substrate \cite{supplementals}.
Scanning tunneling microscopy (STM) was performed at liquid nitrogen temperatures \cite{mun17}.

\subsection*{Theory}
Density functional theory (DFT) calculations were performed on the \SSY\  endofullerene because it is chemically very similar to \SST\ but easier to describe accurately \cite{sta18}.
The calculations on the isolated \SSY\ cluster were performed with ORCA \cite{Neese_2020_a} and the calculations of \SSY/\hBN/Ni(111) using Quantum ESPRESSO \cite{giannozzi09}; details are given in the supplemental material \cite{supplementals}.

We denote the calculated structures $h$-BN/Ni(111)-($\sqrt{19}\times\sqrt{19}$)R$23^\circ$, $h$-BN/Ni(111)-($\sqrt{21}\times\sqrt{21}$)R$11^\circ$ and $h$-BN/Ni(111)-($6\times{}6$) as $\sqrt{19}$, $\sqrt{21}$ and $6\times{}6$ or ``isolated'', respectively. We define the condensation energy $E_c$ as $E_c=E_\text{tot} - E_\text{mol} - E_\text{sub}$, where $E_\text{tot}$ is the total energy of the fully relaxed system, $E_\text{mol}$ the energy of an isolated \YSS\ molecule, and $E_{\text{sub}}$ the energy of the relaxed substrate without molecule. The lock in energy is the energy difference between the highest energy and the lowest energy upon translations of the molecular layer relative to the substrate.
We approximate the ionisation potential and electron affinity by the energy eigenvalues of the highest occupied and lowest unoccupied molecular orbital (HOMO and LUMO), respectively.

%***********************************************************************
\section*{Results and Discussion}

\subsection*{Incommensurability and low temperature super structure}

\begin{figure}
\begin{center}\includegraphics[width=1\columnwidth]{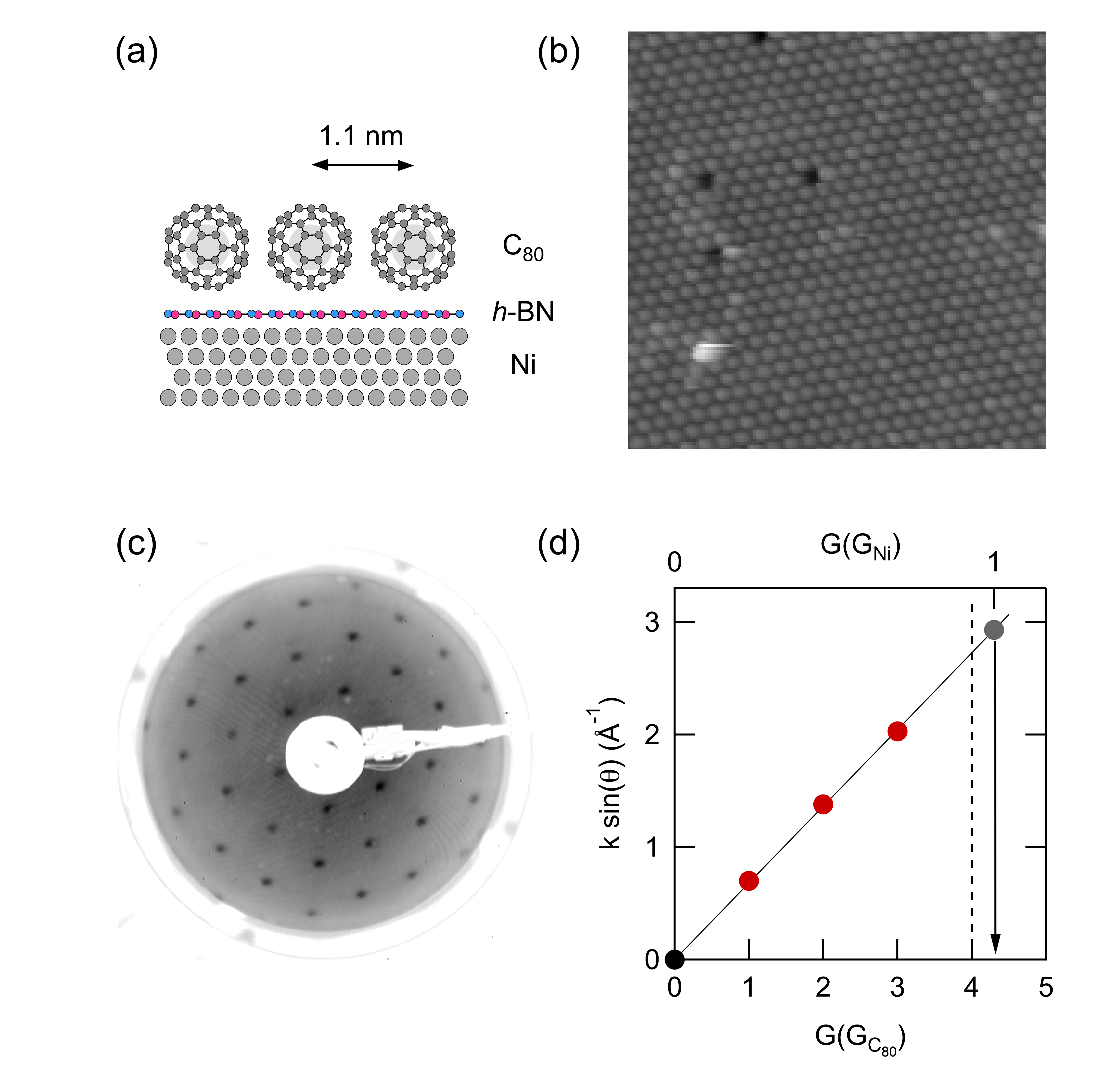}\end{center}
\caption{\SST\ on $h$-BN/Ni(111). (a) Sketch. (b) Tunneling microscopy image of a monolayer preparation recorded at 77~K. (Area = $25 \times 25$~nm$^2$, $I_t$= 1~nA, $V_t$= -5~mV). In the hexagonally close packed layer single vacancies and one second layer molecule are detected. (c) LEED pattern ($E=30$~eV) recorded at room temperature. An ordered C$_{80}$ ($1\times{}1$)  structure with a lattice constant of $1.09\pm 0.05$~nm is discerned. (d) Parallel component of the LEED scattering vector of TbSc$_2$N@C$_{80}$ from panel (c) (red) and $h$-BN/Ni(111) (grey) versus reciprocal lattice vectors G$_{\mathrm{Ni}}$ and G$_{\mathrm{C}_{80}}$. 
%is 4$\pi/\sqrt{3}\, a$, where $a$ is the corresponding lattice constant. 
The C$_{80}$ and the Ni lattices are aligned and incommensurate, where one \Ceig\ fits on $\approx$4.3$\times$4.3 substrate unit cells.}
\label{F1}
\end{figure}

Figure \ref{F1} depicts the investigated system: A monolayer of \SST\ on \hBN/Ni(111). The STM image shows hexagonally close packed molecules, and LEED indicates large scale ordering along the high symmetry directions of the substrate. While \Csix\ on $h$-BN/Ni(111) forms a commensurate Ni (4$\times$4) superstructure \cite{mun05}, the \Ceig\  molecules are about a factor of $\sqrt{80/60}$=1.15 larger and thus may not accommodate in 4$\times$4 Ni(111) unit cells. The analysis of the LEED spot positions suggests an aligned but incommensurate $\approx$4.3$\times$4.3 super structure of  \SST\ on $h$-BN/Ni(111).

\begin{figure}
\begin{center}\includegraphics[width=1\columnwidth]{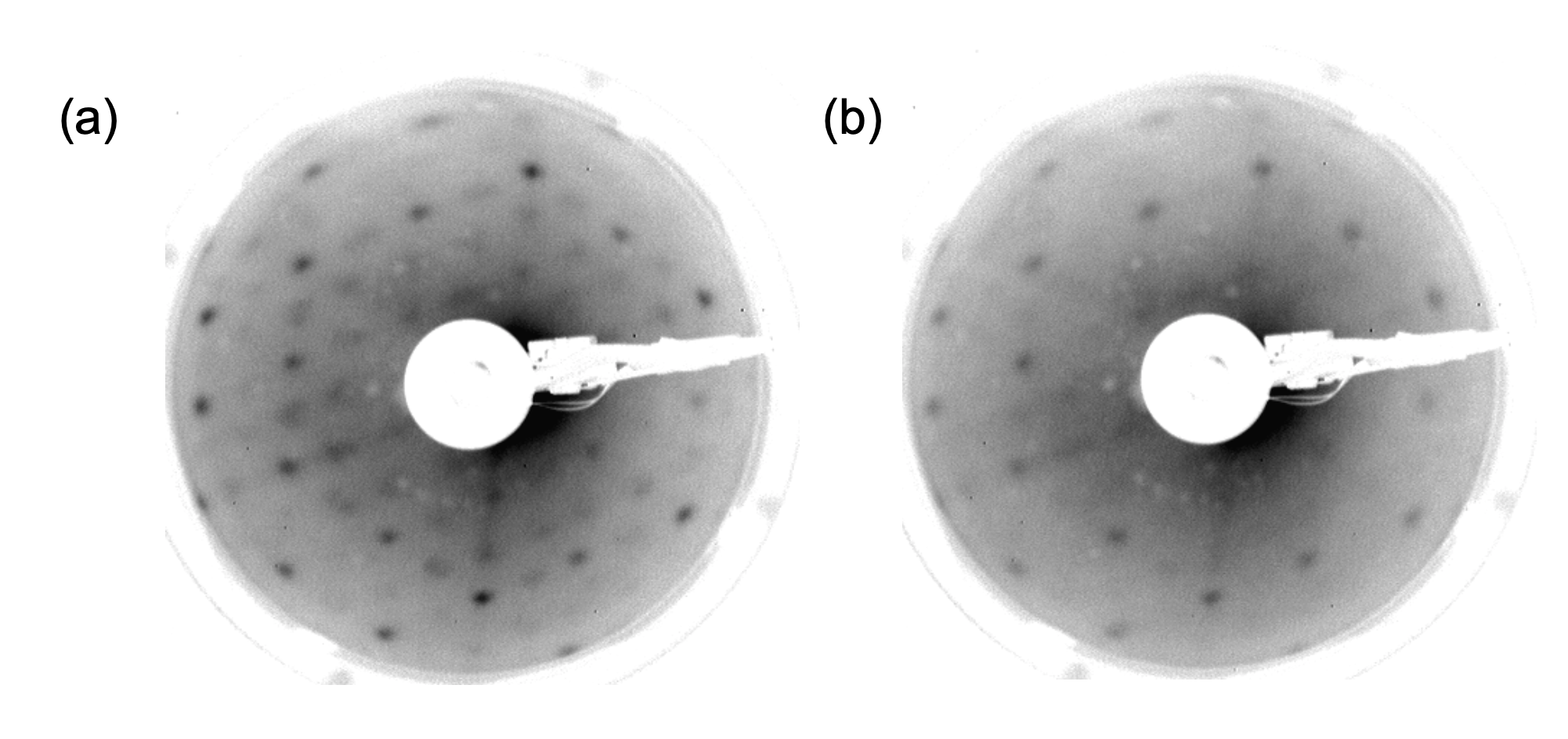}\end{center}
\caption{LEED ($E=26.4$~eV) from \SST\ on {\mbox{$h$-BN/Ni(111)}} at two different temperatures. (a) C$_{80}$ (2$\times$2) phase at 83~K. (b) At 170~K the C$_{80}$ (1$\times$1) phase dominates as at room temperature (see Figure~\ref{F1}(c)). }
\label{F11}
\end{figure}

Figure \ref{F11} shows the formation of a C$_{80}$ (2$\times$2) low temperature superstructure in the incommensurate monolayer \SST\ on \hBN/Ni(111). The structure forms between 80 and 170~K.
Such low temperature ordering or freezing of fullerenes is known from \Csix\ films, where below 100~K a C$_{60}$ (2$\times$2) structure was established \cite{ben93,gol96}. For {\mbox{\Csix\ on \hBN /Ni(111)}} a $(\sqrt{3}\times\sqrt{3})$ phase was found below 160~K \cite{mun05}.
This indicates that like in three dimensions \cite{hei91}, intermolecular forces lead to ordering of the fullerenes below room temperature in two dimensional systems. 
The interaction between \SST\ molecules is, however, expected to be more complicated than that between \Csix\  because the high symmetry of the \Ceig\ cage is broken by the endohedral unit, which is reflected in their appearance as ellipsoids that are weakly distorted from spherical shape and their permanent dipole moments. 

In order to better understand the \SST\ on $h$-BN/Ni(111) system and its differences to {\mbox{\Csix\ on \hBN /Ni(111)}} extensive DFT calculations were performed.
Table~\ref{T1} compares different calculated molecular properties of C$_{60}$ and \YSS, which is, except for the paramagnetism of Tb, electronically very similar to \SST, that was used in the experiments.
\begin{table}[ht]
\centering
\begin{tabular}{ l r |c |c }\hline\hline
 && ~~~~~C$_{60}$~~~~~ & \YSS  \\
\hline
semi minor axis $a$ &(nm) &0.355 &0.417 \\
semi minor axis $b$ &(nm)  &0.355 & 0.411\\
semi minor axis $c$ &(nm)  &0.355&0.409 \\
volume &(nm$^3$)&0.524&  0.723\\
void volume&(nm$^3$)&0.038&0.080\\
dipole moment &(D)&0&0.252\\
electron affinity &(eV) & 4.19 & 3.81 \\
ionization energy &(eV) & 5.83 & 5.32 \\\hline\hline
\end{tabular}
\caption{Calculated geometric and electronic properties of \Csix\ and \YSS\ in the gas phase. The volume and void volume have been calculated in adding or subtracting the van der Waals radius of a carbon atom, 0.145 nm, to the semi minor axes $a$, $b$ and $c$ and then applying the standard formula.}
\label{T1}
\end{table}

\begin{figure}
\begin{center}\includegraphics[width=1\columnwidth]{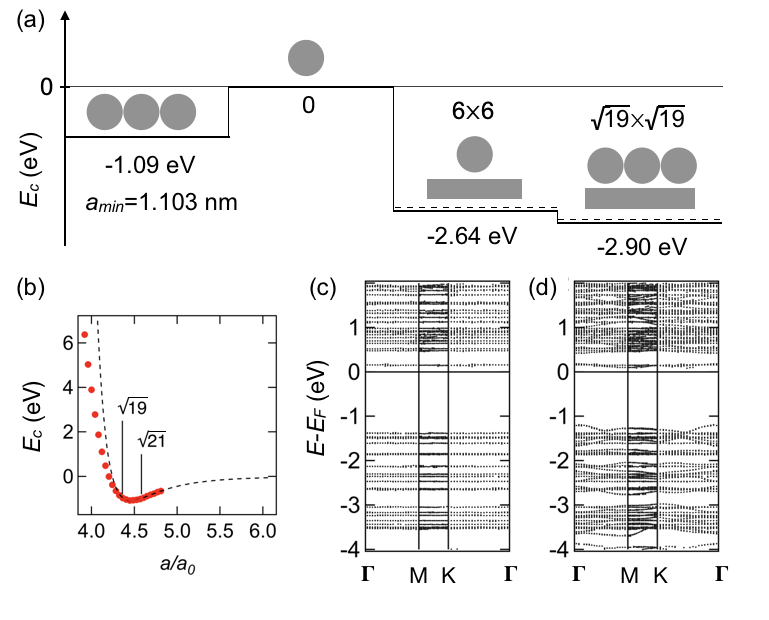}\end{center}
\caption{DFT results on the condensation of \YSS. (a) Condensation energy $E_c$ of \YSS\ in a monolayer (-1.09~eV), free (reference$\equiv$0), isolated on \hBN/Ni(111)\ {\mbox{(-2.64~eV)}}, and in a \hBN/Ni(111)-$\sqrt{19}$ unit cell (-2.90~eV). The dashed lines indicate the lock in energy of the molecules relative to the lowest energy structures.
(b) $E_c$ vs. lattice constant $a$ in a freestanding monolayer \YSS\ in substrate units $a_0$. The lattice constants of the $\sqrt{19}$ and $\sqrt{21}$ super cells are indicated. The dashed curve is the fit of a Lennard-Jones potential at $a >4.36~a_0$ with a radial offset of 2.49~$a_0$. Band structure of the carbon atoms of \YSS\ with respect to the Fermi energy $E_F$ in (c) (6$\times$6) and (d)  $\sqrt{19}$.}
\label{F2}
\end{figure}

Figure \ref{F2} shows the DFT results of a freestanding \YSS\ monolayer, \YSS\ on $h$-BN/Ni(111)-$\sqrt{19}$, and \YSS\ on {\mbox{$h$-BN/Ni(111)-$6\times 6$}}. As a reference we refer to \YSS\ in the vacuum. In Figure~\ref{F2}(a) the condensation energies $E_c$ are displayed.
Figure~\ref{F2}(b) shows the energy of \YSS\ in a two dimensional hexagonally close packed layer. The minimum corresponds to a lattice constant of $a_\text{min} = 1.103$~nm and a condensation energy of -1.09~eV. 
Because the molecules do not fit in a $4 \times4$ Ni(111) unit cell the modelling of the system with DFT is involved.
A commensurate unit cell that is in line with the LEED observation would not be rotated relative to the substrate lattice, and would have to contain ten or more molecules, which can not be handled with present-day capabilities. 
Instead, we found the lock in energy of the molecules to be small and performed DFT calculations in rotated unit cells that contain one molecule. 
The lattice constant of a $\sqrt{19}$
substrate unit cell is close to the experimentally inferred lattice constant and we used it as a commensurate approximation of the  \YSS\ on \hBN/Ni(111) system.
Some calculations were performed in the larger $\sqrt{21}$ cell.
Compared to the free standing layer the $\sqrt{19}$ unit cell causes a compressive strain of 2.6~\% and a 124~meV higher energy,
while in the $\sqrt{21}$ an expansive strain of 2.4~\% and a 88~meV higher energy was found.
Calculations where the lateral position of the molecule was shifted and fixed relative to the substrate yielded lock in energies of 99 and 90~meV for the $\sqrt{19}$ and the $6\times 6$ structures. 
If the molecules are incommensurate with the substrate the average energy per molecule increases by 49~meV and 34~meV, respectively \cite{supplementals}.
Therefore, the lock in energy is small compared to thermal energies at the preparation temperature, the binding energy in the freestanding layer, and the adsorption energy of a single \YSS\ on $h$-BN/Ni(111). This supports the picture of non-covalent bonding of the molecules to the surface and easy incommensurate layer formation.
The orientation of the incommensurate molecular lattice is likely guided by atomic steps in the substrate that run along high symmetry $\langle1 {\bar{1}}0 \rangle$ directions.   

In Figures \ref{F2}(c) and (d) the band structure of the valence- and conduction-bands on the carbon cages of \YSS\ on $h$-BN/Ni(111) in $6\times{}6$ and $\sqrt{19}$ cells are shown.
While the bands in the former "isolated" structure are flat, the dispersion in the latter, condensed phase is considerable and around the Fermi level of $p$-type, dispersing downward, with maxima at the $\Gamma$ point.
The large HOMO--LUMO gaps of the molecules become apparent (1.49 and 1.28 eV), where the LUMO is pinned 118 and 81 meV above the Fermi level in the two structures, respectively.
For comparison the HOMO--LUMO gap of the free molecule is 1.37~eV.
We note that the applied exchange-correlation functional does not match with the experimental gap inferred from optical absorption spectroscopy, where 1.72~eV for \YSS\ in toluene solution %723 nm
has been measured \cite{chen07}.

Endofullerenes are known to have conformers with local energy minima that are difficult to find with DFT optimization methods.
For the search of the lowest energy structure we used an approach where the optimisations start from 120 different conformations \cite{dub19}. 
As for C$_{80}$ endofullerenes in vacuum  \cite{yan08,dub19,Westerstroem21} and on surfaces \cite{dub19} we found different conformers depending on the conformation at the start of the optimization \cite{supplementals}. 
The calculated workfunction in the lowest energy $\sqrt{19}$ structure is 4.30~eV, while it is 3.55~eV for the substrate without molecules. 
The standard deviation of the workfunctions of the 120 optimizations is 79~meV and that of the corresponding condensation energies is 68~meV  \cite{supplementals}.
Within the performed simulations with one molecule per unit cell we found no correlation between workfunction and condensation energy. At this point we speculate that the ($2 \times 2$) molecular ordering, which we could not afford to calculate, might reduce the workfunction by a non-parallel arrangement of the molecular dipoles.

In order to cross check the DFT model and the experimental structures we calculated N~1s photoemission core level binding energies for hexagonal boron nitride on Ni(111), with and without molecules, and the endohedral nitrogen in \YSS\ on \hBN/Ni(111)-$\sqrt{19}$. 

\begin{figure}
\begin{center}
\includegraphics[width=1\columnwidth]{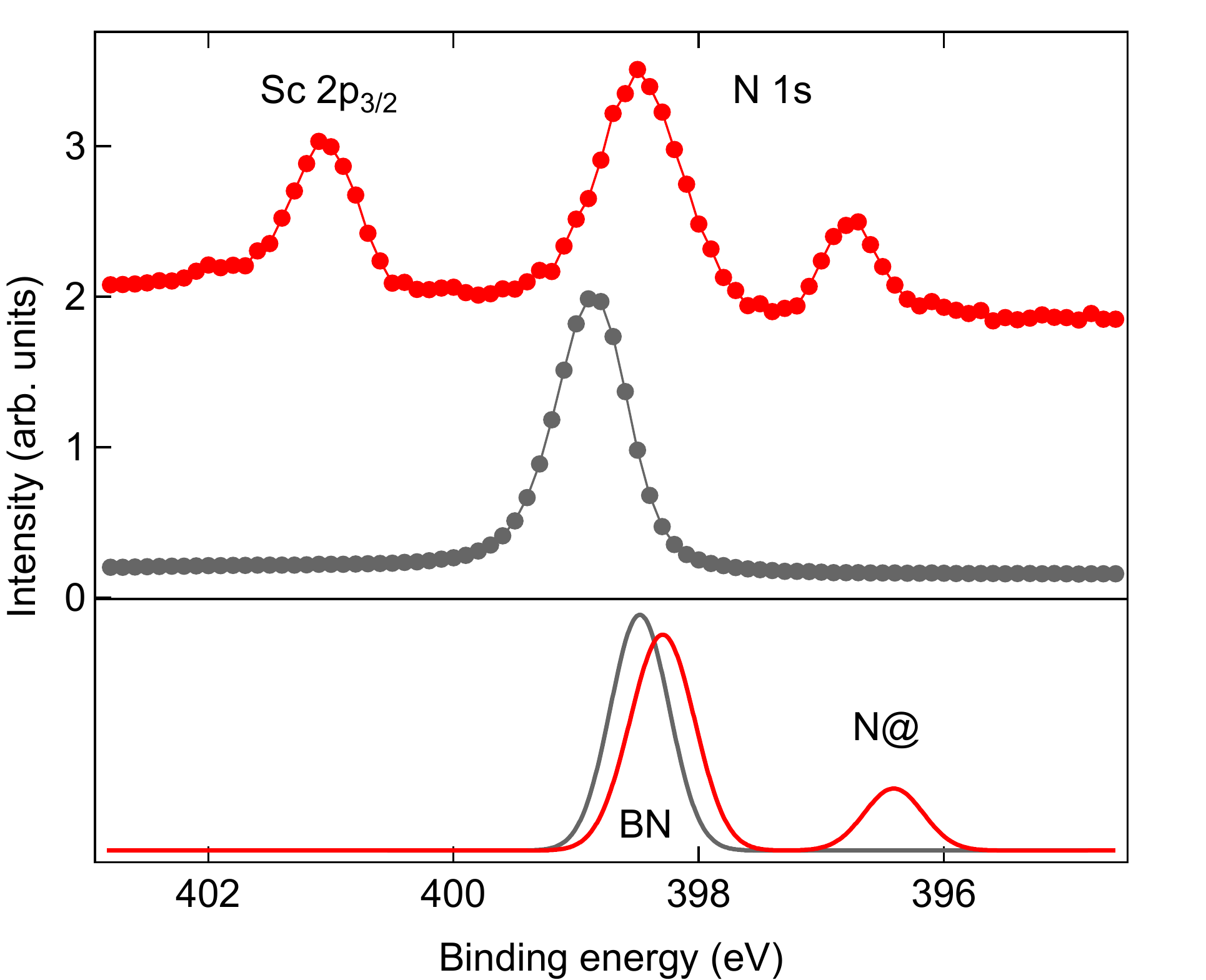}
\end{center}
\caption{Comparison of XPS and DFT N~1s energies. Top panel, XPS on \hBN/Ni(111) (grey dots) and \SST\ on \hBN/Ni(111) (red dots) at room temperature (photon energy $h\nu$=600~eV). Bottom panel, DFT final state eigenvalues convoluted with a Gaussian with 470~meV FWHM for \hBN/Ni(111) (dark grey line)  \YSS\ on {\mbox{\hBN/Ni(111) $\sqrt{19}$}} (red line).  
The weight of the BN is the same for  \hBN/Ni(111) and  \YSS\ on \hBN/Ni(111) and that of the  endohedral nitrogen is increased by a factor of 5. 
The  N 1s final state energy in \YSS\ from DFT is matched to the experimental binding energy in \SST \cite{supplementals}. }
\label{F22}
\end{figure}

In Figure~\ref{F22} x-ray photoelectron spectroscopy (XPS) data are compared with final state energies that are convoluted with a Gaussian of 470 meV full width at half maximum (FWHM).
Inspection increases the confidence that experiment and theory describe very similar physical situations. 
In contrast to initial state core level calculations of the C$_{80}$ cage \cite{sta18}, the final state had to be considered \cite{supplementals}. 
Self-consistent calculations with a half hole on the N~1s orbital were performed and the resulting eigenvalues were used. 
From these calculations we reached an excellent agreement of the lowest energy $\sqrt{19}$ structure and the experiment (see Figure~\ref{F22} and the Table in the supplemental material \cite{supplementals}). 
The DFT results confirm the assignment of the N~1s core level peaks to the boron nitride and to the endohedral nitrogen species \cite{sta18}. The broadening of the BN derived N~1s peak upon adsorption of endohedral C$_{80}$ predicts that the nitrogen atoms in the BN layer corrugate under the influence of the adsorbed molecules to 0.09~nm \cite{supplementals}. 
The shift of the BN derived N~1s peak upon adsorption of C$_{80}$ is parallel to the work function and is in line with the physisorption picture of \hBN\ on Ni that involves vacuum level alignment \cite{nag95}. 
The chemical shift of the endohedral N~1s peak agrees for experiments at room temperature, but, as we will see below, not for the low temperature case, where ($2 \times 2$)  molecular ordering was observed.

\begin{figure}
\begin{center}\includegraphics[width=1\columnwidth]{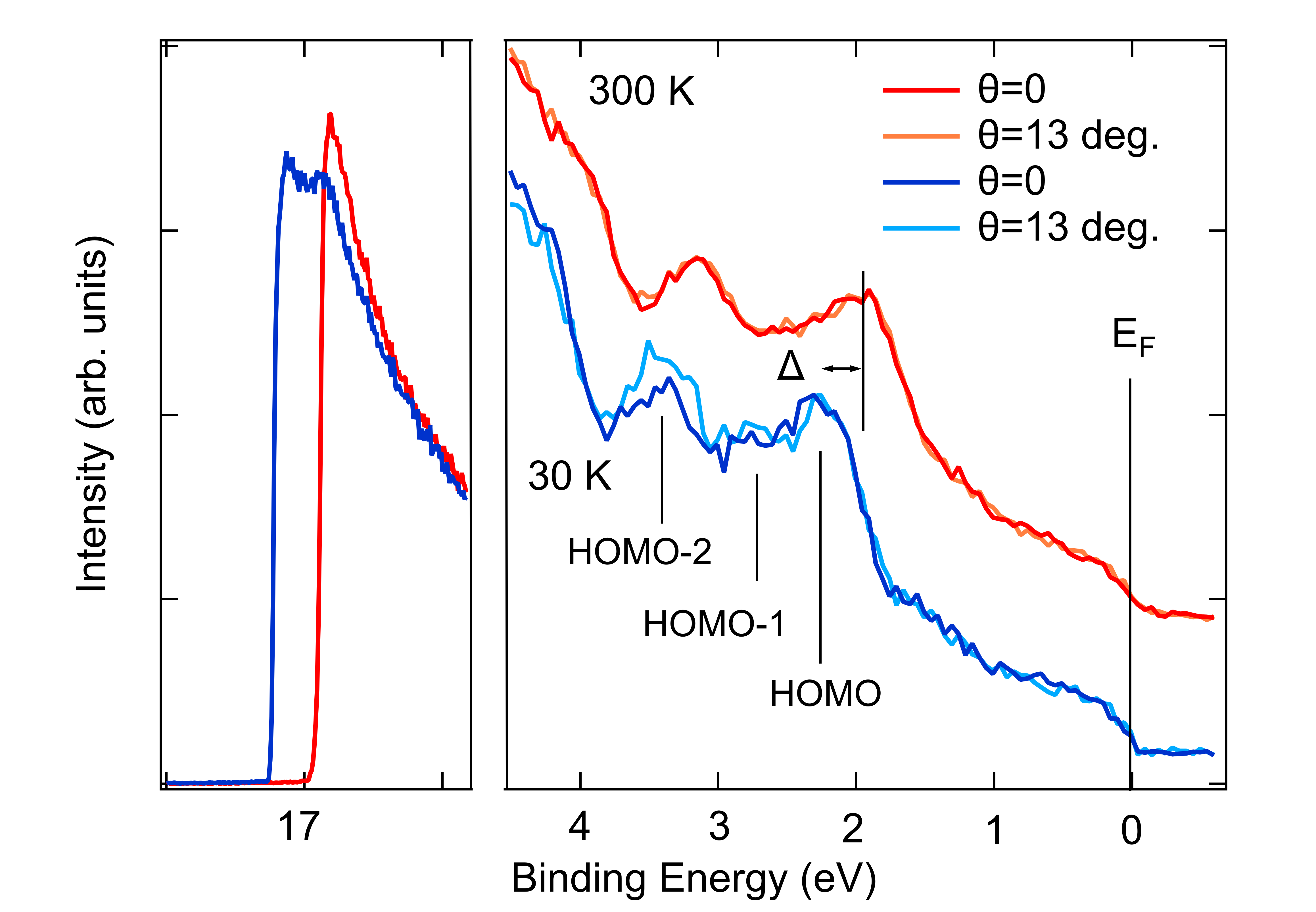}\end{center}
\caption{Angular resolved He~I$\alpha$ excited valence band photoemission (ARUPS) from \SST\ on \hBN /Ni(111). The temperature induced shift $\Delta= 0.3$~eV of the molecular orbital-peaks labeled as HOMO, HOMO-1 and HOMO-2 is parallel to the shift of the secondary electron cut off (left panel) that indicates a workfunction increase from 3.95~eV at 30~K to 4.20~eV at 300~K. At low temperature the molecular orbitals display anisotropy with polar emission angle $\theta$, while at room temperature they do not.}
\label{F3}
\end{figure}

Figure~\ref{F3} displays He~I$\alpha$ excited valence band photoemission data and the corresponding secondary electron cutoff of \SST\ on $h$-BN/Ni(111) at 30 and 300~K. The workfunction increases with temperature from 3.95 to 4.20~eV. This shift is also reflected in the energy of the HOMO and it indicates vacuum level alignment of the \SST\ molecular orbitals, which is in line with a non-covalent bonding to the substrate.
The HOMO peak of \SST\ on {\mbox{$h$-BN/Ni(111)}} occurs at relatively high binding energies of 2.30 and 2.04~eV at 30 and 300~K. They are larger than the optical HOMO--LUMO gap of \YSS\ and indicative for a strong on site Coulomb interaction U \cite{lof92,rud99}. In multilayer \SSS\ \cite{alv02} the HOMO was found at $\approx$1.73~eV binding energy.
In contrast to {\mbox{\Csix\ on $h$-BN/Ni(111)}} \cite{mun05} we find no hint on temperature induced charge transfer, namely no partial filling of the LUMO that was seen as extra intensity at the Fermi level. The valence band data show, however, anisotropy of the molecular orbitals at low temperature, while no anisotropy is visible at room temperature. Such intensity variations are known as ultra-violet photoelectron diffraction (UPD) effects \cite{ost96,pus09,gre13} and indicate orientational order of the molecules at low temperatures.

\subsection*{Behaviour of the molecular conformation with temperature}

\begin{figure}[t]
\begin{center}\includegraphics[width=1\columnwidth]{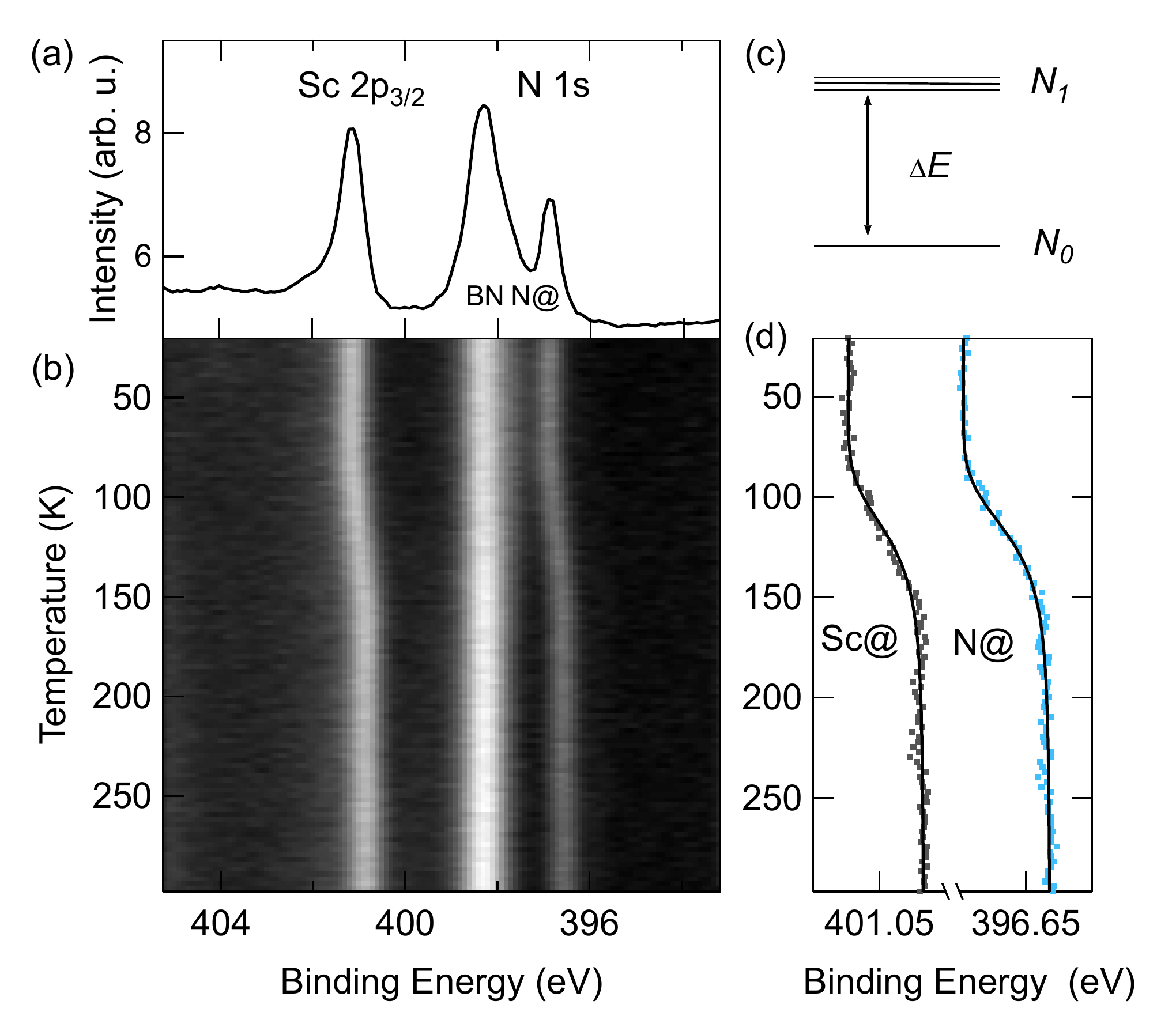}\end{center}
\caption{XPS from \SST\ on \hBN /Ni(111) (photon energy $h\nu$=600~eV). (a) Spectrum in the binding energy range of the N~1s and the Sc~2p$_{3/2}$ core levels. At low temperature the endohedral nitrogen species has a 1.42~eV lower binding energy than the nitrogen of \hBN. (b) Temperature dependent XPS between 30~ and 300~K. (c) Two level model for the temperature dependence of the molecular core level energy shifts. The energy difference $\Delta E$ and the degeneracy $N_1$ are the only thermodynamic parameters. (d) Endohedral Sc~2p$_{3/2}$ (grey) and N~1s (light blue) binding energies as a function of the sample temperature with a fit of the two level model -- solid lines.}
\label{F4}
\end{figure}

Figure~\ref{F4} shows XPS of \SST\ on {\mbox{\hBN /Ni(111)}} between 30 K and room temperature. The N~1s and the Sc~3p$_{3/2}$ core levels are observed. The  N~1s level of the endohedral unit displays relative to the N~1s levels of the BN substrate sharper and at lower binding energy. 
%The  1.43 and 1.73$\pm0.05$ eV lower binding energy. The molecule based N$^{3-}$ species displays sharper and at lower binding energy (see Figure \ref{F22}). 
The BN N~1s orbitals of the substrate have constant binding energy ($\pm$25~meV), while those of the atoms in the molecules follow the workfunction, i.e. show the same energy shift as the secondary cut off and the HOMO molecular orbitals in Figure \ref{F3}. This means that the BN N~1s energy does not follow the temperature dependence of the vacuum level, while the one of the molecular layer on top of \hBN\ does.

We assign the observed behaviour of the workfunction to the onset of endohedral rotation and different dipole components $p_z$ along the surface normal. 
The workfunction change $\Delta \Phi$ and $\Delta p_z$ are connected via the Helmholtz equation $\Delta \Phi=-n \Delta p_z e/\epsilon_0$, where $n$ is the areal density of the corresponding dipole component $\Delta p_z$ and $e$ the elementary charge.
Taking the calculated gas phase dipole moment of \YSS\ of 0.25~D we would get a workfunction increase of 92 meV in a $\sqrt{19}$ unit cell in going from a configuration where the dipole points antiparallel to the surface normal to an isotropic configuration with a zero average dipole moment. 

The temperature dependence of the core level binding energies that are parallel to the workfunction allows to extract thermodynamic properties of the \SST\ on {\mbox{\hBN /Ni(111)}} phase transition.
Specifically, we consider for the molecular conformations two energy levels that are separated by $\Delta E=E_1-E_0$. While $E_0$ is set to be non-degenerate ($N_0=1$), the degeneracy of $E_1$ is $N_1$
(see Figure \ref{F4}(c)).
As it can be seen from the calculations in the supplementary material \cite{supplementals}, the assumption of two conformational energies only is a simplification, though the "smooth bend" of the order parameter does not allow for reliably fitting of more than four independent parameters.  
At thermal equilibrium we obtain the occupation $\langle n_0\rangle$ of the lowest energy state $E_0$,
\begin{equation}
\langle n_0\rangle=\frac{1}{1+N_1\exp{(-\Delta E/k_BT})}
\label{E2}
\end{equation}
and the occupation of state $E_1$, $\langle n_1\rangle=1-\langle n_0\rangle$.

If the measured core level energies $E_B$ are taken to express the order parameter i.e. the weighted sum {\mbox{$E_B= \langle n_0\rangle E_{B0}+\langle n_1\rangle E_{B1}$}}, we may derive $E_{B0}$, $E_{B1}$, $\Delta E$ and $N_1$ from the measured data. $E_{B0}$ describes the system at T=0 or $\langle n_1\rangle=0$  and $E_{B1}$ is the core level energy if all molecules were excited, i.e. $\langle n_1\rangle=1$. 
%, where $E_{B0}$ and $E_{B1}$ are the core level binding energies of the two respective states in the model. 
From  $\langle n_0\rangle (T)$  the relevant thermodynamic quantities $\Delta E$ and $N_1$ are determined. 
The solid lines in Figure \ref{F4}(d) show the fits of the $N_1$ degenerate two level model for the N~1s and the Sc~2p$_{3/2}$ core levels of \SST\ on \hBN /Ni(111). The fit-parameters are $\Delta E= 74\pm6$~meV, $E_{B0}-E_{B1} = 310\pm40$~meV  and the degeneracy $N_1=1300\pm700$. 
As expected $E_{B0}-E_{B1}$ is in line with the work function shift. The energy $\Delta E$ has the order of magnitude of the scatter of $E_c$ for the calculated conformations of \YSS\ on \hBN/Ni(111) \cite{supplementals}. This means that the transition has the energy scale of different molecular conformations of \SSY /\hBN /Ni(111).
The large degeneracy $N_1$ reflects the large C$_{80}$ ($2 \times 2$) unit cell at low temperature and indicates a strong contribution of the entropy to the free energy of the system above the ordering temperature. 

\section*{Conclusions}

We observed a temperature induced workfunction shift of \SST\ on \hBN /Ni(111) centered at 125 K.
This shift is related to the disappearance of a \SST (2$\times 2$) LEED structure.
The transition is an order disorder transition with onset of rotation of the endohedral clusters.
This conclusion bases on angular dependent valence band photoelectron spectroscopy.
DFT calculations, predict the lattice constant of \SSY , a permanent dipole moment, describe the N~1s XPS spectrum in the \SSY (1$\times 1$) phase and propose different conformers with an energy distribution that is in line with a model for the temperature dependence of the molecular core level binding energies.
%With the presented findings we put forward the converse effect i.e. that the molecular conformation may be controlled with an electric field as found in the surface dipole that establishes the workfunction. 
%By this means the magnetism of the the endohedral units could be remotely controlled with electric fields.


\subsection*{Acknowledgments}
This work was supported by the Swiss National Science Foundation grant No.\ 200020\_153312 and the Institute for Basic Science (IBS-R014-D1).
Calculations were performed at the CSCS Swiss National Supercomputing Centre under project uzh11.
The photoemission, LEED and STM measurements were performed at the PEARL beamline of the Swiss Light Source, Paul Scherrer Institut, Villigen. AAP acknowledges Deutsche Forschungsgemeinschaft (grant PO 1602/8-1).



\begin{thebibliography}{36}%
\makeatletter
\providecommand \@ifxundefined [1]{%
 \@ifx{#1\undefined}
}%
\providecommand \@ifnum [1]{%
 \ifnum #1\expandafter \@firstoftwo
 \else \expandafter \@secondoftwo
 \fi
}%
\providecommand \@ifx [1]{%
 \ifx #1\expandafter \@firstoftwo
 \else \expandafter \@secondoftwo
 \fi
}%
\providecommand \natexlab [1]{#1}%
\providecommand \enquote  [1]{``#1''}%
\providecommand \bibnamefont  [1]{#1}%
\providecommand \bibfnamefont [1]{#1}%
\providecommand \citenamefont [1]{#1}%
\providecommand \href@noop [0]{\@secondoftwo}%
\providecommand \href [0]{\begingroup \@sanitize@url \@href}%
\providecommand \@href[1]{\@@startlink{#1}\@@href}%
\providecommand \@@href[1]{\endgroup#1\@@endlink}%
\providecommand \@sanitize@url [0]{\catcode `\\12\catcode `\$12\catcode
  `\&12\catcode `\#12\catcode `\^12\catcode `\_12\catcode `\%12\relax}%
\providecommand \@@startlink[1]{}%
\providecommand \@@endlink[0]{}%
\providecommand \url  [0]{\begingroup\@sanitize@url \@url }%
\providecommand \@url [1]{\endgroup\@href {#1}{\urlprefix }}%
\providecommand \urlprefix  [0]{URL }%
\providecommand \Eprint [0]{\href }%
\providecommand \doibase [0]{http://dx.doi.org/}%
\providecommand \selectlanguage [0]{\@gobble}%
\providecommand \bibinfo  [0]{\@secondoftwo}%
\providecommand \bibfield  [0]{\@secondoftwo}%
\providecommand \translation [1]{[#1]}%
\providecommand \BibitemOpen [0]{}%
\providecommand \bibitemStop [0]{}%
\providecommand \bibitemNoStop [0]{.\EOS\space}%
\providecommand \EOS [0]{\spacefactor3000\relax}%
\providecommand \BibitemShut  [1]{\csname bibitem#1\endcsname}%
\let\auto@bib@innerbib\@empty
%</preamble>
\bibitem [{\citenamefont {Rudolf}\ \emph {et~al.}(1999)\citenamefont {Rudolf},
  \citenamefont {Golden},\ and\ \citenamefont {Br\"uhwiler}}]{rud99}%
  \BibitemOpen
  \bibfield  {author} {\bibinfo {author} {\bibfnamefont {P.}~\bibnamefont
  {Rudolf}}, \bibinfo {author} {\bibfnamefont {M.~S.}\ \bibnamefont {Golden}},
  \ and\ \bibinfo {author} {\bibfnamefont {P.~A}\ \bibnamefont {Br\"uhwiler}},\
  }\bibfield  {title} {\enquote {\bibinfo {title} {Studies of fullerenes by the
  excitation, emission, and scattering of electrons},}\ }\href {\doibase
  https://doi.org/10.1016/S0368-2048(99)00058-4} {\bibfield  {journal}
  {\bibinfo  {journal} {J. Electron Spectros. Relat. Phenomena}\ }\textbf
  {\bibinfo {volume} {100}},\ \bibinfo {pages} {409--433} (\bibinfo {year}
  {1999})}\BibitemShut {NoStop}%
\bibitem [{\citenamefont {S\'anchez}\ \emph {et~al.}(2009)\citenamefont
  {S\'anchez}, \citenamefont {Otero}, \citenamefont {Gallego}, \citenamefont
  {Miranda},\ and\ \citenamefont {Martin}}]{san09}%
  \BibitemOpen
  \bibfield  {author} {\bibinfo {author} {\bibfnamefont {L.}~\bibnamefont
  {S\'anchez}}, \bibinfo {author} {\bibfnamefont {R.}~\bibnamefont {Otero}},
  \bibinfo {author} {\bibfnamefont {J.~M.}\ \bibnamefont {Gallego}}, \bibinfo
  {author} {\bibfnamefont {R.}~\bibnamefont {Miranda}}, \ and\ \bibinfo
  {author} {\bibfnamefont {N.}~\bibnamefont {Martin}},\ }\bibfield  {title}
  {\enquote {\bibinfo {title} {Ordering fullerenes at the nanometer scale on
  solid surfaces},}\ }\href {\doibase 10.1021/cr800441b} {\bibfield  {journal}
  {\bibinfo  {journal} {Chem. Rev.}\ }\textbf {\bibinfo {volume} {109}},\
  \bibinfo {pages} {2081--2091} (\bibinfo {year} {2009})}\BibitemShut {NoStop}%
\bibitem [{\citenamefont {Moriarty}(2010)}]{mor101}%
  \BibitemOpen
  \bibfield  {author} {\bibinfo {author} {\bibfnamefont {P.~J.}\ \bibnamefont
  {Moriarty}},\ }\bibfield  {title} {\enquote {\bibinfo {title} {Fullerene
  adsorption on semiconductor surfaces},}\ }\href {\doibase
  https://doi.org/10.1016/j.surfrep.2010.08.001} {\bibfield  {journal}
  {\bibinfo  {journal} {Surf. Sci. Rep.}\ }\textbf {\bibinfo {volume} {65}},\
  \bibinfo {pages} {175--227} (\bibinfo {year} {2010})}\BibitemShut {NoStop}%
\bibitem [{\citenamefont {Heiney}\ \emph {et~al.}(1991)\citenamefont {Heiney},
  \citenamefont {Fischer}, \citenamefont {McGhie}, \citenamefont {Romanow},
  \citenamefont {Denenstein}, \citenamefont {McCauley~Jr.}, \citenamefont
  {Smith},\ and\ \citenamefont {Cox}}]{hei91}%
  \BibitemOpen
  \bibfield  {author} {\bibinfo {author} {\bibfnamefont {P.~A.}\ \bibnamefont
  {Heiney}}, \bibinfo {author} {\bibfnamefont {J.~E.}\ \bibnamefont {Fischer}},
  \bibinfo {author} {\bibfnamefont {A.~R.}\ \bibnamefont {McGhie}}, \bibinfo
  {author} {\bibfnamefont {W.~J.}\ \bibnamefont {Romanow}}, \bibinfo {author}
  {\bibfnamefont {A.~M.}\ \bibnamefont {Denenstein}}, \bibinfo {author}
  {\bibfnamefont {J.~P.}\ \bibnamefont {McCauley~Jr.}}, \bibinfo {author}
  {\bibfnamefont {A.~B.}\ \bibnamefont {Smith}}, \ and\ \bibinfo {author}
  {\bibfnamefont {D.~E.}\ \bibnamefont {Cox}},\ }\bibfield  {title} {\enquote
  {\bibinfo {title} {Orientational ordering transition in solid {C}$_{60}$},}\
  }\href {\doibase 10.1103/PhysRevLett.66.2911} {\bibfield  {journal} {\bibinfo
   {journal} {Phys. Rev. Lett.}\ }\textbf {\bibinfo {volume} {66}},\ \bibinfo
  {pages} {2911--2914} (\bibinfo {year} {1991})}\BibitemShut {NoStop}%
\bibitem [{\citenamefont {Benning}\ \emph {et~al.}(1993)\citenamefont
  {Benning}, \citenamefont {Stepniak},\ and\ \citenamefont {Weaver}}]{ben93}%
  \BibitemOpen
  \bibfield  {author} {\bibinfo {author} {\bibfnamefont {P.~J.}\ \bibnamefont
  {Benning}}, \bibinfo {author} {\bibfnamefont {F.}~\bibnamefont {Stepniak}}, \
  and\ \bibinfo {author} {\bibfnamefont {J.~H.}\ \bibnamefont {Weaver}},\
  }\bibfield  {title} {\enquote {\bibinfo {title} {Electron-diffraction and
  photoelectron-spectroscopy studies of fullerene and alkali-metal fulleride
  films},}\ }\href {\doibase 10.1103/PhysRevB.48.9086} {\bibfield  {journal}
  {\bibinfo  {journal} {Phys. Rev. B}\ }\textbf {\bibinfo {volume} {48}},\
  \bibinfo {pages} {9086--9096} (\bibinfo {year} {1993})}\BibitemShut {NoStop}%
\bibitem [{\citenamefont {Goldoni}\ \emph {et~al.}(1996)\citenamefont
  {Goldoni}, \citenamefont {Cepek},\ and\ \citenamefont {Modesti}}]{gol96}%
  \BibitemOpen
  \bibfield  {author} {\bibinfo {author} {\bibfnamefont {A.}~\bibnamefont
  {Goldoni}}, \bibinfo {author} {\bibfnamefont {C.}~\bibnamefont {Cepek}}, \
  and\ \bibinfo {author} {\bibfnamefont {S.}~\bibnamefont {Modesti}},\
  }\bibfield  {title} {\enquote {\bibinfo {title} {First-order
  orientational-disordering transition on the (111) surface of {C}$_{60}$},}\
  }\href {\doibase 10.1103/PhysRevB.54.2890} {\bibfield  {journal} {\bibinfo
  {journal} {Phys. Rev. B}\ }\textbf {\bibinfo {volume} {54}},\ \bibinfo
  {pages} {2890--2895} (\bibinfo {year} {1996})}\BibitemShut {NoStop}%
\bibitem [{\citenamefont {Goldoni}\ \emph {et~al.}(2002)\citenamefont
  {Goldoni}, \citenamefont {Cepek}, \citenamefont {Larciprete}, \citenamefont
  {Sangaletti}, \citenamefont {Pagliara}, \citenamefont {Paolucci},\ and\
  \citenamefont {Sancrotti}}]{gol02}%
  \BibitemOpen
  \bibfield  {author} {\bibinfo {author} {\bibfnamefont {A.}~\bibnamefont
  {Goldoni}}, \bibinfo {author} {\bibfnamefont {C.}~\bibnamefont {Cepek}},
  \bibinfo {author} {\bibfnamefont {R.}~\bibnamefont {Larciprete}}, \bibinfo
  {author} {\bibfnamefont {L.}~\bibnamefont {Sangaletti}}, \bibinfo {author}
  {\bibfnamefont {S.}~\bibnamefont {Pagliara}}, \bibinfo {author}
  {\bibfnamefont {G.}~\bibnamefont {Paolucci}}, \ and\ \bibinfo {author}
  {\bibfnamefont {M.}~\bibnamefont {Sancrotti}},\ }\bibfield  {title} {\enquote
  {\bibinfo {title} {Core level photoemission evidence of frustrated surface
  molecules: A germ of disorder at the (111) surface of {C}$_{60}$ before the
  order-disorder surface phase transition},}\ }\href {\doibase
  10.1103/PhysRevLett.88.196102} {\bibfield  {journal} {\bibinfo  {journal}
  {Phys. Rev. Lett.}\ }\textbf {\bibinfo {volume} {88}},\ \bibinfo {pages}
  {196102} (\bibinfo {year} {2002})}\BibitemShut {NoStop}%
\bibitem [{\citenamefont {Muntwiler}\ \emph {et~al.}(2005)\citenamefont
  {Muntwiler}, \citenamefont {Auw{\"a}rter}, \citenamefont {Seitsonen},
  \citenamefont {Osterwalder},\ and\ \citenamefont {Greber}}]{mun05}%
  \BibitemOpen
  \bibfield  {author} {\bibinfo {author} {\bibfnamefont {M.}~\bibnamefont
  {Muntwiler}}, \bibinfo {author} {\bibfnamefont {W.}~\bibnamefont
  {Auw{\"a}rter}}, \bibinfo {author} {\bibfnamefont {A.}~\bibnamefont
  {Seitsonen}}, \bibinfo {author} {\bibfnamefont {J.}~\bibnamefont
  {Osterwalder}}, \ and\ \bibinfo {author} {\bibfnamefont {T.}~\bibnamefont
  {Greber}},\ }\bibfield  {title} {\enquote {\bibinfo {title}
  {Rocking-motion-induced charging of {C}$_{60}$ on $h$-{BN/Ni}(111)},}\ }\href
  {\doibase 10.1103/PhysRevB.71.121402} {\bibfield  {journal} {\bibinfo
  {journal} {Phys. Rev. B}\ }\textbf {\bibinfo {volume} {71}},\ \bibinfo
  {pages} {121402} (\bibinfo {year} {2005})}\BibitemShut {NoStop}%
\bibitem [{\citenamefont {Popov}\ \emph {et~al.}({2013})\citenamefont {Popov},
  \citenamefont {Yang},\ and\ \citenamefont {Dunsch}}]{pop13}%
  \BibitemOpen
  \bibfield  {author} {\bibinfo {author} {\bibfnamefont {A.~A.}\ \bibnamefont
  {Popov}}, \bibinfo {author} {\bibfnamefont {S.}~\bibnamefont {Yang}}, \ and\
  \bibinfo {author} {\bibfnamefont {L.}~\bibnamefont {Dunsch}},\ }\bibfield
  {title} {\enquote {\bibinfo {title} {{Endohedral Fullerenes}},}\ }\href
  {\doibase {10.1021/cr300297r}} {\bibfield  {journal} {\bibinfo  {journal}
  {{Chem. Rev.}}\ }\textbf {\bibinfo {volume} {{113}}},\ \bibinfo {pages}
  {{5989--6113}} (\bibinfo {year} {{2013}})}\BibitemShut {NoStop}%
\bibitem [{\citenamefont {Stevenson}\ \emph {et~al.}(1999)\citenamefont
  {Stevenson}, \citenamefont {Rice}, \citenamefont {Glass}, \citenamefont
  {Harich}, \citenamefont {Cromer}, \citenamefont {Jordan}, \citenamefont
  {Craft}, \citenamefont {Hadju}, \citenamefont {Bible}, \citenamefont
  {Olmstead}, \citenamefont {Maitra}, \citenamefont {Fisher}, \citenamefont
  {Balch},\ and\ \citenamefont {Dorn}}]{ste99}%
  \BibitemOpen
  \bibfield  {author} {\bibinfo {author} {\bibfnamefont {S.}~\bibnamefont
  {Stevenson}}, \bibinfo {author} {\bibfnamefont {G.}~\bibnamefont {Rice}},
  \bibinfo {author} {\bibfnamefont {T.}~\bibnamefont {Glass}}, \bibinfo
  {author} {\bibfnamefont {K.}~\bibnamefont {Harich}}, \bibinfo {author}
  {\bibfnamefont {F.}~\bibnamefont {Cromer}}, \bibinfo {author} {\bibfnamefont
  {M.~R.}\ \bibnamefont {Jordan}}, \bibinfo {author} {\bibfnamefont
  {J.}~\bibnamefont {Craft}}, \bibinfo {author} {\bibfnamefont
  {E.}~\bibnamefont {Hadju}}, \bibinfo {author} {\bibfnamefont
  {R.}~\bibnamefont {Bible}}, \bibinfo {author} {\bibfnamefont {M.~M.}\
  \bibnamefont {Olmstead}}, \bibinfo {author} {\bibfnamefont {K.}~\bibnamefont
  {Maitra}}, \bibinfo {author} {\bibfnamefont {A.~J.}\ \bibnamefont {Fisher}},
  \bibinfo {author} {\bibfnamefont {A.~L.}\ \bibnamefont {Balch}}, \ and\
  \bibinfo {author} {\bibfnamefont {H.~C.}\ \bibnamefont {Dorn}},\ }\bibfield
  {title} {\enquote {\bibinfo {title} {Small-bandgap endohedral
  metallofullerenes in high yield and purity},}\ }\href {\doibase
  10.1038/43415} {\bibfield  {journal} {\bibinfo  {journal} {Nature}\ }\textbf
  {\bibinfo {volume} {401}},\ \bibinfo {pages} {55--57} (\bibinfo {year}
  {1999})}\BibitemShut {NoStop}%
\bibitem [{\citenamefont {Westerstr\"om}\ \emph {et~al.}({2012})\citenamefont
  {Westerstr\"om}, \citenamefont {Dreiser}, \citenamefont {Piamonteze},
  \citenamefont {Muntwiler}, \citenamefont {Weyeneth}, \citenamefont {Brune},
  \citenamefont {Rusponi}, \citenamefont {Nolting}, \citenamefont {Popov},
  \citenamefont {Yang}, \citenamefont {Dunsch},\ and\ \citenamefont
  {Greber}}]{wes12}%
  \BibitemOpen
  \bibfield  {author} {\bibinfo {author} {\bibfnamefont {R.}~\bibnamefont
  {Westerstr\"om}}, \bibinfo {author} {\bibfnamefont {J.}~\bibnamefont
  {Dreiser}}, \bibinfo {author} {\bibfnamefont {C.}~\bibnamefont {Piamonteze}},
  \bibinfo {author} {\bibfnamefont {M.}~\bibnamefont {Muntwiler}}, \bibinfo
  {author} {\bibfnamefont {S.}~\bibnamefont {Weyeneth}}, \bibinfo {author}
  {\bibfnamefont {H.}~\bibnamefont {Brune}}, \bibinfo {author} {\bibfnamefont
  {S.}~\bibnamefont {Rusponi}}, \bibinfo {author} {\bibfnamefont
  {F.}~\bibnamefont {Nolting}}, \bibinfo {author} {\bibfnamefont
  {A.}~\bibnamefont {Popov}}, \bibinfo {author} {\bibfnamefont {SF.}\
  \bibnamefont {Yang}}, \bibinfo {author} {\bibfnamefont {L.}~\bibnamefont
  {Dunsch}}, \ and\ \bibinfo {author} {\bibfnamefont {T.}~\bibnamefont
  {Greber}},\ }\bibfield  {title} {\enquote {\bibinfo {title} {An endohedral
  single-molecule magnet with long relaxation times: {DySc$_2$N@C$_{80}$}},}\
  }\href {\doibase {10.1021/ja301044p}} {\bibfield  {journal} {\bibinfo
  {journal} {{J. Am. Chem. Soc.}}\ }\textbf {\bibinfo {volume} {{134}}},\
  \bibinfo {pages} {{9840--9843}} (\bibinfo {year} {{2012}})}\BibitemShut
  {NoStop}%
\bibitem [{\citenamefont {Alvarez}\ \emph {et~al.}(2002)\citenamefont
  {Alvarez}, \citenamefont {Pichler}, \citenamefont {Georgi}, \citenamefont
  {Schwieger}, \citenamefont {Peisert}, \citenamefont {Dunsch}, \citenamefont
  {Hu}, \citenamefont {Knupfer}, \citenamefont {Fink}, \citenamefont
  {Bressler}, \citenamefont {Mast},\ and\ \citenamefont {Golden}}]{alv02}%
  \BibitemOpen
  \bibfield  {author} {\bibinfo {author} {\bibfnamefont {L.}~\bibnamefont
  {Alvarez}}, \bibinfo {author} {\bibfnamefont {T.}~\bibnamefont {Pichler}},
  \bibinfo {author} {\bibfnamefont {P.}~\bibnamefont {Georgi}}, \bibinfo
  {author} {\bibfnamefont {T.}~\bibnamefont {Schwieger}}, \bibinfo {author}
  {\bibfnamefont {H.}~\bibnamefont {Peisert}}, \bibinfo {author} {\bibfnamefont
  {L.}~\bibnamefont {Dunsch}}, \bibinfo {author} {\bibfnamefont
  {Z.}~\bibnamefont {Hu}}, \bibinfo {author} {\bibfnamefont {M.}~\bibnamefont
  {Knupfer}}, \bibinfo {author} {\bibfnamefont {J.}~\bibnamefont {Fink}},
  \bibinfo {author} {\bibfnamefont {P.}~\bibnamefont {Bressler}}, \bibinfo
  {author} {\bibfnamefont {M.}~\bibnamefont {Mast}}, \ and\ \bibinfo {author}
  {\bibfnamefont {M.~S.}\ \bibnamefont {Golden}},\ }\bibfield  {title}
  {\enquote {\bibinfo {title} {Electronic structure of pristine and
  intercalated {Sc$_{3}$N@C$_{80}$} metallofullerene},}\ }\href {\doibase
  10.1103/PhysRevB.66.035107} {\bibfield  {journal} {\bibinfo  {journal} {Phys.
  Rev. B}\ }\textbf {\bibinfo {volume} {66}},\ \bibinfo {pages} {035107}
  (\bibinfo {year} {2002})}\BibitemShut {NoStop}%
\bibitem [{\citenamefont {Shiozawa}\ \emph {et~al.}(2005)\citenamefont
  {Shiozawa}, \citenamefont {Rauf}, \citenamefont {Pichler}, \citenamefont
  {Grimm}, \citenamefont {Liu}, \citenamefont {Knupfer}, \citenamefont
  {Kalbac}, \citenamefont {Yang}, \citenamefont {Dunsch}, \citenamefont
  {B\"uchner},\ and\ \citenamefont {Batchelor}}]{shi05}%
  \BibitemOpen
  \bibfield  {author} {\bibinfo {author} {\bibfnamefont {H.}~\bibnamefont
  {Shiozawa}}, \bibinfo {author} {\bibfnamefont {H.}~\bibnamefont {Rauf}},
  \bibinfo {author} {\bibfnamefont {T.}~\bibnamefont {Pichler}}, \bibinfo
  {author} {\bibfnamefont {D.}~\bibnamefont {Grimm}}, \bibinfo {author}
  {\bibfnamefont {X.}~\bibnamefont {Liu}}, \bibinfo {author} {\bibfnamefont
  {M.}~\bibnamefont {Knupfer}}, \bibinfo {author} {\bibfnamefont
  {M.}~\bibnamefont {Kalbac}}, \bibinfo {author} {\bibfnamefont
  {S.}~\bibnamefont {Yang}}, \bibinfo {author} {\bibfnamefont {L.}~\bibnamefont
  {Dunsch}}, \bibinfo {author} {\bibfnamefont {B.}~\bibnamefont {B\"uchner}}, \
  and\ \bibinfo {author} {\bibfnamefont {D.}~\bibnamefont {Batchelor}},\
  }\bibfield  {title} {\enquote {\bibinfo {title} {Electronic structure of the
  trimetal nitride fullerene {Dy$_{3}$N}@{C}$_{80}$},}\ }\href {\doibase
  10.1103/PhysRevB.72.195409} {\bibfield  {journal} {\bibinfo  {journal} {Phys.
  Rev. B}\ }\textbf {\bibinfo {volume} {72}},\ \bibinfo {pages} {195409}
  (\bibinfo {year} {2005})}\BibitemShut {NoStop}%
\bibitem [{\citenamefont {Leigh}\ \emph {et~al.}({2007})\citenamefont {Leigh},
  \citenamefont {Norenberg}, \citenamefont {Cattaneo}, \citenamefont {Owen},
  \citenamefont {Porfyrakis}, \citenamefont {Bassi}, \citenamefont {Ardavan},\
  and\ \citenamefont {Briggs}}]{lei07}%
  \BibitemOpen
  \bibfield  {author} {\bibinfo {author} {\bibfnamefont {D.~F.}\ \bibnamefont
  {Leigh}}, \bibinfo {author} {\bibfnamefont {C.}~\bibnamefont {Norenberg}},
  \bibinfo {author} {\bibfnamefont {D.}~\bibnamefont {Cattaneo}}, \bibinfo
  {author} {\bibfnamefont {J.~H.~G.}\ \bibnamefont {Owen}}, \bibinfo {author}
  {\bibfnamefont {K.}~\bibnamefont {Porfyrakis}}, \bibinfo {author}
  {\bibfnamefont {A.~Li}\ \bibnamefont {Bassi}}, \bibinfo {author}
  {\bibfnamefont {A.}~\bibnamefont {Ardavan}}, \ and\ \bibinfo {author}
  {\bibfnamefont {G.~A.~D.}\ \bibnamefont {Briggs}},\ }\bibfield  {title}
  {\enquote {\bibinfo {title} {{Self-assembly of trimetallic nitride template
  fullerenes on surfaces studied by STM}},}\ }\href {\doibase
  {10.1016/j.susc.2006.12.035}} {\bibfield  {journal} {\bibinfo  {journal}
  {{Surf. Sci.}}\ }\textbf {\bibinfo {volume} {{601}}},\ \bibinfo {pages}
  {{2750--2755}} (\bibinfo {year} {{2007}})}\BibitemShut {NoStop}%
\bibitem [{\citenamefont {Treier}\ \emph {et~al.}(2009)\citenamefont {Treier},
  \citenamefont {Ruffieux}, \citenamefont {Fasel}, \citenamefont {Nolting},
  \citenamefont {Yang}, \citenamefont {Dunsch},\ and\ \citenamefont
  {Greber}}]{tre09}%
  \BibitemOpen
  \bibfield  {author} {\bibinfo {author} {\bibfnamefont {M.}~\bibnamefont
  {Treier}}, \bibinfo {author} {\bibfnamefont {P.}~\bibnamefont {Ruffieux}},
  \bibinfo {author} {\bibfnamefont {R.}~\bibnamefont {Fasel}}, \bibinfo
  {author} {\bibfnamefont {F.}~\bibnamefont {Nolting}}, \bibinfo {author}
  {\bibfnamefont {SF.}\ \bibnamefont {Yang}}, \bibinfo {author} {\bibfnamefont
  {L.}~\bibnamefont {Dunsch}}, \ and\ \bibinfo {author} {\bibfnamefont
  {T.}~\bibnamefont {Greber}},\ }\bibfield  {title} {\enquote {\bibinfo {title}
  {Looking inside an endohedral fullerene: Inter- and intramolecular ordering
  of {${\text{Dy}}_{3}\text{N}@{\text{C}}_{80}$ $({I}_{h})$} on {Cu(111)}},}\
  }\href {\doibase 10.1103/PhysRevB.80.081403} {\bibfield  {journal} {\bibinfo
  {journal} {Phys. Rev. B}\ }\textbf {\bibinfo {volume} {80}},\ \bibinfo
  {pages} {081403} (\bibinfo {year} {2009})}\BibitemShut {NoStop}%
\bibitem [{\citenamefont {Huang}\ \emph {et~al.}({2011})\citenamefont {Huang},
  \citenamefont {Zhao}, \citenamefont {Peng}, \citenamefont {Popov},
  \citenamefont {Yang}, \citenamefont {Dunsch},\ and\ \citenamefont
  {Petek}}]{tia11}%
  \BibitemOpen
  \bibfield  {author} {\bibinfo {author} {\bibfnamefont {T.}~\bibnamefont
  {Huang}}, \bibinfo {author} {\bibfnamefont {J.}~\bibnamefont {Zhao}},
  \bibinfo {author} {\bibfnamefont {M.}~\bibnamefont {Peng}}, \bibinfo {author}
  {\bibfnamefont {A.~A.}\ \bibnamefont {Popov}}, \bibinfo {author}
  {\bibfnamefont {SF.}\ \bibnamefont {Yang}}, \bibinfo {author} {\bibfnamefont
  {L.}~\bibnamefont {Dunsch}}, \ and\ \bibinfo {author} {\bibfnamefont
  {H.}~\bibnamefont {Petek}},\ }\bibfield  {title} {\enquote {\bibinfo {title}
  {A molecular switch based on current-driven rotation of an encapsulated
  cluster within a fullerene cage},}\ }\href {\doibase {10.1021/nl2028409}}
  {\bibfield  {journal} {\bibinfo  {journal} {{Nano Lett.}}\ }\textbf {\bibinfo
  {volume} {{11}}},\ \bibinfo {pages} {{5327--5332}} (\bibinfo {year}
  {{2011}})}\BibitemShut {NoStop}%
\bibitem [{\citenamefont {Krylov}\ \emph {et~al.}({2020})\citenamefont
  {Krylov}, \citenamefont {Schimmel}, \citenamefont {Dubrovin}, \citenamefont
  {Liu}, \citenamefont {Nhung~Nguyen}, \citenamefont {Spree}, \citenamefont
  {Chen}, \citenamefont {Velkos}, \citenamefont {Bulbucan}, \citenamefont
  {Westerstr\"om}, \citenamefont {Studniarek}, \citenamefont {Dreiser},
  \citenamefont {Hess}, \citenamefont {B\"uchner}, \citenamefont {Avdoshenko},\
  and\ \citenamefont {Popov}}]{kry20}%
  \BibitemOpen
  \bibfield  {author} {\bibinfo {author} {\bibfnamefont {D.~S.}\ \bibnamefont
  {Krylov}}, \bibinfo {author} {\bibfnamefont {S.}~\bibnamefont {Schimmel}},
  \bibinfo {author} {\bibfnamefont {V.}~\bibnamefont {Dubrovin}}, \bibinfo
  {author} {\bibfnamefont {F.}~\bibnamefont {Liu}}, \bibinfo {author}
  {\bibfnamefont {T.~T.}\ \bibnamefont {Nhung~Nguyen}}, \bibinfo {author}
  {\bibfnamefont {L.}~\bibnamefont {Spree}}, \bibinfo {author} {\bibfnamefont
  {C.H.}\ \bibnamefont {Chen}}, \bibinfo {author} {\bibfnamefont
  {G.}~\bibnamefont {Velkos}}, \bibinfo {author} {\bibfnamefont
  {C.}~\bibnamefont {Bulbucan}}, \bibinfo {author} {\bibfnamefont
  {R.}~\bibnamefont {Westerstr\"om}}, \bibinfo {author} {\bibfnamefont
  {M.}~\bibnamefont {Studniarek}}, \bibinfo {author} {\bibfnamefont
  {J.}~\bibnamefont {Dreiser}}, \bibinfo {author} {\bibfnamefont
  {C.}~\bibnamefont {Hess}}, \bibinfo {author} {\bibfnamefont {B.}~\bibnamefont
  {B\"uchner}}, \bibinfo {author} {\bibfnamefont {S.~M.}\ \bibnamefont
  {Avdoshenko}}, \ and\ \bibinfo {author} {\bibfnamefont {A.~A.}\ \bibnamefont
  {Popov}},\ }\bibfield  {title} {\enquote {\bibinfo {title}
  {Substrate-independent magnetic bistability in monolayers of the
  single-molecule magnet {D}y$_2${S}c{N@}{C}$_{80}$ on metals and
  insulators},}\ }\href {\doibase {10.1002/anie.201913955}} {\bibfield
  {journal} {\bibinfo  {journal} {{Angew. Chem. Int. Ed.}}\ }\textbf {\bibinfo
  {volume} {{59}}},\ \bibinfo {pages} {{5756--5764}} (\bibinfo {year}
  {{2020}})}\BibitemShut {NoStop}%
\bibitem [{\citenamefont {Westerstr\"om}\ \emph {et~al.}({2015})\citenamefont
  {Westerstr\"om}, \citenamefont {Uldry}, \citenamefont {Stania}, \citenamefont
  {Dreiser}, \citenamefont {Piamonteze}, \citenamefont {Muntwiler},
  \citenamefont {Matsui}, \citenamefont {Rusponi}, \citenamefont {Brune},
  \citenamefont {Yang}, \citenamefont {Popov}, \citenamefont {B\"uchner},
  \citenamefont {Delley},\ and\ \citenamefont {Greber}}]{wes15}%
  \BibitemOpen
  \bibfield  {author} {\bibinfo {author} {\bibfnamefont {R.}~\bibnamefont
  {Westerstr\"om}}, \bibinfo {author} {\bibfnamefont {A.-C.}\ \bibnamefont
  {Uldry}}, \bibinfo {author} {\bibfnamefont {R.}~\bibnamefont {Stania}},
  \bibinfo {author} {\bibfnamefont {J.}~\bibnamefont {Dreiser}}, \bibinfo
  {author} {\bibfnamefont {C.}~\bibnamefont {Piamonteze}}, \bibinfo {author}
  {\bibfnamefont {M.}~\bibnamefont {Muntwiler}}, \bibinfo {author}
  {\bibfnamefont {F.}~\bibnamefont {Matsui}}, \bibinfo {author} {\bibfnamefont
  {S.}~\bibnamefont {Rusponi}}, \bibinfo {author} {\bibfnamefont
  {H.}~\bibnamefont {Brune}}, \bibinfo {author} {\bibfnamefont {SF.}\
  \bibnamefont {Yang}}, \bibinfo {author} {\bibfnamefont {A.}~\bibnamefont
  {Popov}}, \bibinfo {author} {\bibfnamefont {B.}~\bibnamefont {B\"uchner}},
  \bibinfo {author} {\bibfnamefont {B.}~\bibnamefont {Delley}}, \ and\ \bibinfo
  {author} {\bibfnamefont {T.}~\bibnamefont {Greber}},\ }\bibfield  {title}
  {\enquote {\bibinfo {title} {Surface aligned magnetic moments and hysteresis
  of an endohedral single-molecule magnet on a metal},}\ }\href {\doibase
  {10.1103/PhysRevLett.114.087201}} {\bibfield  {journal} {\bibinfo  {journal}
  {{Phys. Rev. Lett.}}\ }\textbf {\bibinfo {volume} {{114}}},\ \bibinfo {pages}
  {{087201--5}} (\bibinfo {year} {{2015}})}\BibitemShut {NoStop}%
\bibitem [{\citenamefont {Greber}\ \emph {et~al.}({2019})\citenamefont
  {Greber}, \citenamefont {Seitsonen}, \citenamefont {Hemmi}, \citenamefont
  {Dreiser}, \citenamefont {Stania}, \citenamefont {Matsui}, \citenamefont
  {Muntwiler}, \citenamefont {Popov},\ and\ \citenamefont
  {Westerstr\"om}}]{gre19}%
  \BibitemOpen
  \bibfield  {author} {\bibinfo {author} {\bibfnamefont {T.}~\bibnamefont
  {Greber}}, \bibinfo {author} {\bibfnamefont {A.~P.}\ \bibnamefont
  {Seitsonen}}, \bibinfo {author} {\bibfnamefont {A.}~\bibnamefont {Hemmi}},
  \bibinfo {author} {\bibfnamefont {J.}~\bibnamefont {Dreiser}}, \bibinfo
  {author} {\bibfnamefont {R.}~\bibnamefont {Stania}}, \bibinfo {author}
  {\bibfnamefont {F.}~\bibnamefont {Matsui}}, \bibinfo {author} {\bibfnamefont
  {M.}~\bibnamefont {Muntwiler}}, \bibinfo {author} {\bibfnamefont {A.~A.}\
  \bibnamefont {Popov}}, \ and\ \bibinfo {author} {\bibfnamefont
  {R.}~\bibnamefont {Westerstr\"om}},\ }\bibfield  {title} {\enquote {\bibinfo
  {title} {{Circular dichroism and angular deviation in x-ray absorption
  spectra of Dy$_2$ScN@C$_{80}$ single-molecule magnets on h-BN/Rh(111)}},}\
  }\href {\doibase {10.1103/PhysRevMaterials.3.014409}} {\bibfield  {journal}
  {\bibinfo  {journal} {{Phys. Rev. Materials}}\ }\textbf {\bibinfo {volume}
  {{3}}},\ \bibinfo {pages} {{014409}} (\bibinfo {year} {{2019}})}\BibitemShut
  {NoStop}%
\bibitem [{\citenamefont {Kostanyan}\ \emph {et~al.}({2017})\citenamefont
  {Kostanyan}, \citenamefont {Westerstr\"om}, \citenamefont {Zhang},
  \citenamefont {Kunhardt}, \citenamefont {Stania}, \citenamefont {B\"uchner},
  \citenamefont {Popov},\ and\ \citenamefont {Greber}}]{kos17}%
  \BibitemOpen
  \bibfield  {author} {\bibinfo {author} {\bibfnamefont {A.}~\bibnamefont
  {Kostanyan}}, \bibinfo {author} {\bibfnamefont {R.}~\bibnamefont
  {Westerstr\"om}}, \bibinfo {author} {\bibfnamefont {Y.}~\bibnamefont
  {Zhang}}, \bibinfo {author} {\bibfnamefont {D.}~\bibnamefont {Kunhardt}},
  \bibinfo {author} {\bibfnamefont {R.}~\bibnamefont {Stania}}, \bibinfo
  {author} {\bibfnamefont {B.}~\bibnamefont {B\"uchner}}, \bibinfo {author}
  {\bibfnamefont {A.~A.}\ \bibnamefont {Popov}}, \ and\ \bibinfo {author}
  {\bibfnamefont {T.}~\bibnamefont {Greber}},\ }\bibfield  {title} {\enquote
  {\bibinfo {title} {Switching molecular conformation with the torque on a
  single magnetic moment},}\ }\href {\doibase {10.1103/PhysRevLett.119.237202}}
  {\bibfield  {journal} {\bibinfo  {journal} {{Phys. Rev. Lett.}}\ }\textbf
  {\bibinfo {volume} {{119}}},\ \bibinfo {pages} {{237202}} (\bibinfo {year}
  {{2017}})}\BibitemShut {NoStop}%
\bibitem [{\citenamefont {Stania}\ \emph {et~al.}({2018})\citenamefont
  {Stania}, \citenamefont {Seitsonen}, \citenamefont {Kunhardt}, \citenamefont
  {B\"uchner}, \citenamefont {Popov}, \citenamefont {Muntwiler},\ and\
  \citenamefont {Greber}}]{sta18}%
  \BibitemOpen
  \bibfield  {author} {\bibinfo {author} {\bibfnamefont {R.}~\bibnamefont
  {Stania}}, \bibinfo {author} {\bibfnamefont {A.~P.}\ \bibnamefont
  {Seitsonen}}, \bibinfo {author} {\bibfnamefont {D.}~\bibnamefont {Kunhardt}},
  \bibinfo {author} {\bibfnamefont {B.}~\bibnamefont {B\"uchner}}, \bibinfo
  {author} {\bibfnamefont {A.~A.}\ \bibnamefont {Popov}}, \bibinfo {author}
  {\bibfnamefont {M.}~\bibnamefont {Muntwiler}}, \ and\ \bibinfo {author}
  {\bibfnamefont {T.}~\bibnamefont {Greber}},\ }\bibfield  {title} {\enquote
  {\bibinfo {title} {Electrostatic interaction across a single-layer carbon
  shell},}\ }\href {\doibase {10.1021/acs.jpclett.8b01326}} {\bibfield
  {journal} {\bibinfo  {journal} {{J. Phys. Chem. Lett.}}\ }\textbf {\bibinfo
  {volume} {{9}}},\ \bibinfo {pages} {{3586--3590}} (\bibinfo {year}
  {{2018}})}\BibitemShut {NoStop}%
\bibitem [{\citenamefont {Zhang}\ \emph {et~al.}(2015)\citenamefont {Zhang},
  \citenamefont {Krylov}, \citenamefont {Rosenkranz}, \citenamefont
  {Schiemenz},\ and\ \citenamefont {Popov}}]{zha15}%
  \BibitemOpen
  \bibfield  {author} {\bibinfo {author} {\bibfnamefont {Y.}~\bibnamefont
  {Zhang}}, \bibinfo {author} {\bibfnamefont {D.}~\bibnamefont {Krylov}},
  \bibinfo {author} {\bibfnamefont {M.}~\bibnamefont {Rosenkranz}}, \bibinfo
  {author} {\bibfnamefont {S.}~\bibnamefont {Schiemenz}}, \ and\ \bibinfo
  {author} {\bibfnamefont {A.~A.}\ \bibnamefont {Popov}},\ }\bibfield  {title}
  {\enquote {\bibinfo {title} {Magnetic anisotropy of endohedral lanthanide
  ions: paramagnetic {NMR} study of {MS}c$_2${N}@{C}$_{80}$-{I}$_h$ with {M}
  running through the whole $4f$ row},}\ }\href {\doibase 10.1039/C5SC00154D}
  {\bibfield  {journal} {\bibinfo  {journal} {Chem. Sci.}\ }\textbf {\bibinfo
  {volume} {6}},\ \bibinfo {pages} {2328--2341} (\bibinfo {year}
  {2015})}\BibitemShut {NoStop}%
\bibitem [{\citenamefont {Auw\"arter}\ \emph {et~al.}(1999)\citenamefont
  {Auw\"arter}, \citenamefont {Kreutz}, \citenamefont {Greber},\ and\
  \citenamefont {Osterwalder}}]{auw99}%
  \BibitemOpen
  \bibfield  {author} {\bibinfo {author} {\bibfnamefont {W.}~\bibnamefont
  {Auw\"arter}}, \bibinfo {author} {\bibfnamefont {T.~J.}\ \bibnamefont
  {Kreutz}}, \bibinfo {author} {\bibfnamefont {T.}~\bibnamefont {Greber}}, \
  and\ \bibinfo {author} {\bibfnamefont {J.}~\bibnamefont {Osterwalder}},\
  }\bibfield  {title} {\enquote {\bibinfo {title} {{XPD} and {STM}
  investigation of hexagonal boron nitride on {N}i(111)},}\ }\href {\doibase
  10.1016/S0039-6028(99)00381-7} {\bibfield  {journal} {\bibinfo  {journal}
  {Surf. Sci.}\ }\textbf {\bibinfo {volume} {429}},\ \bibinfo {pages}
  {229--236} (\bibinfo {year} {1999})}\BibitemShut {NoStop}%
\bibitem [{\citenamefont {Muntwiler}\ \emph {et~al.}({2017})\citenamefont
  {Muntwiler}, \citenamefont {Zhang}, \citenamefont {Stania}, \citenamefont
  {Matsui}, \citenamefont {Oberta}, \citenamefont {Flechsig}, \citenamefont
  {Patthey}, \citenamefont {Quitmann}, \citenamefont {Glatzel}, \citenamefont
  {Widmer}, \citenamefont {Meyer}, \citenamefont {Jung}, \citenamefont {Aebi},
  \citenamefont {Fasel},\ and\ \citenamefont {Greber}}]{mun17}%
  \BibitemOpen
  \bibfield  {author} {\bibinfo {author} {\bibfnamefont {M.}~\bibnamefont
  {Muntwiler}}, \bibinfo {author} {\bibfnamefont {J.}~\bibnamefont {Zhang}},
  \bibinfo {author} {\bibfnamefont {R.}~\bibnamefont {Stania}}, \bibinfo
  {author} {\bibfnamefont {F.}~\bibnamefont {Matsui}}, \bibinfo {author}
  {\bibfnamefont {P.}~\bibnamefont {Oberta}}, \bibinfo {author} {\bibfnamefont
  {U.}~\bibnamefont {Flechsig}}, \bibinfo {author} {\bibfnamefont
  {L.}~\bibnamefont {Patthey}}, \bibinfo {author} {\bibfnamefont
  {C.}~\bibnamefont {Quitmann}}, \bibinfo {author} {\bibfnamefont
  {T.}~\bibnamefont {Glatzel}}, \bibinfo {author} {\bibfnamefont
  {R.}~\bibnamefont {Widmer}}, \bibinfo {author} {\bibfnamefont
  {E.}~\bibnamefont {Meyer}}, \bibinfo {author} {\bibfnamefont {T.~A.}\
  \bibnamefont {Jung}}, \bibinfo {author} {\bibfnamefont {P.}~\bibnamefont
  {Aebi}}, \bibinfo {author} {\bibfnamefont {R.}~\bibnamefont {Fasel}}, \ and\
  \bibinfo {author} {\bibfnamefont {T.}~\bibnamefont {Greber}},\ }\bibfield
  {title} {\enquote {\bibinfo {title} {Surface science at the {PEARL} beamline
  of the {S}wiss {L}ight {S}ource},}\ }\href {\doibase
  {10.1107/S1600577516018646}} {\bibfield  {journal} {\bibinfo  {journal} {{J.
  Synchrotron Rad.}}\ }\textbf {\bibinfo {volume} {{24}}},\ \bibinfo {pages}
  {{354--366}} (\bibinfo {year} {{2017}})}\BibitemShut {NoStop}%
\bibitem [{sup()}]{supplementals}%
  \BibitemOpen
  \href@noop {} {\enquote {\bibinfo {title} {Supplementaries},}\ }\BibitemShut
  {NoStop}%
\bibitem [{\citenamefont {Neese}\ \emph {et~al.}(2020)\citenamefont {Neese},
  \citenamefont {Wennmohs}, \citenamefont {Becker},\ and\ \citenamefont
  {Riplinger}}]{Neese_2020_a}%
  \BibitemOpen
  \bibfield  {author} {\bibinfo {author} {\bibfnamefont {F.}~\bibnamefont
  {Neese}}, \bibinfo {author} {\bibfnamefont {F.}~\bibnamefont {Wennmohs}},
  \bibinfo {author} {\bibfnamefont {U.}~\bibnamefont {Becker}}, \ and\ \bibinfo
  {author} {\bibfnamefont {C.}~\bibnamefont {Riplinger}},\ }\bibfield  {title}
  {\enquote {\bibinfo {title} {The {ORCA} quantum chemistry program package},}\
  }\href {\doibase 10.1063/5.0004608} {\bibfield  {journal} {\bibinfo
  {journal} {J. Chem. Phys.}\ }\textbf {\bibinfo {volume} {152}},\ \bibinfo
  {pages} {224108} (\bibinfo {year} {2020})}\BibitemShut {NoStop}%
\bibitem [{\citenamefont {Giannozzi}\ \emph {et~al.}(2009)\citenamefont
  {Giannozzi}, \citenamefont {Baroni}, \citenamefont {Bonini}, \citenamefont
  {Calandra}, \citenamefont {Car}, \citenamefont {Cavazzoni}, \citenamefont
  {Ceresoli}, \citenamefont {Chiarotti}, \citenamefont {Cococcioni},
  \citenamefont {Dabo}, \citenamefont {{Dal Corso}}, \citenamefont {{de
  Gironcoli}}, \citenamefont {Fabris}, \citenamefont {Fratesi}, \citenamefont
  {Gebauer}, \citenamefont {Gerstmann}, \citenamefont {Gougoussis},
  \citenamefont {Kokalj}, \citenamefont {Lazzeri}, \citenamefont
  {Martin-Samos}, \citenamefont {Marzari}, \citenamefont {Mauri}, \citenamefont
  {Mazzarello}, \citenamefont {Paolini}, \citenamefont {Pasquarello},
  \citenamefont {Paulatto}, \citenamefont {Sbraccia}, \citenamefont {Scandolo},
  \citenamefont {Sclauzero}, \citenamefont {Seitsonen}, \citenamefont
  {Smogunov}, \citenamefont {Umari},\ and\ \citenamefont
  {Wentzcovitch}}]{giannozzi09}%
  \BibitemOpen
  \bibfield  {author} {\bibinfo {author} {\bibfnamefont {P.}~\bibnamefont
  {Giannozzi}}, \bibinfo {author} {\bibfnamefont {S.}~\bibnamefont {Baroni}},
  \bibinfo {author} {\bibfnamefont {N.}~\bibnamefont {Bonini}}, \bibinfo
  {author} {\bibfnamefont {M.}~\bibnamefont {Calandra}}, \bibinfo {author}
  {\bibfnamefont {R.}~\bibnamefont {Car}}, \bibinfo {author} {\bibfnamefont
  {C.}~\bibnamefont {Cavazzoni}}, \bibinfo {author} {\bibfnamefont
  {D.}~\bibnamefont {Ceresoli}}, \bibinfo {author} {\bibfnamefont {G.~L.}\
  \bibnamefont {Chiarotti}}, \bibinfo {author} {\bibfnamefont {M.}~\bibnamefont
  {Cococcioni}}, \bibinfo {author} {\bibfnamefont {I.}~\bibnamefont {Dabo}},
  \bibinfo {author} {\bibfnamefont {A.}~\bibnamefont {{Dal Corso}}}, \bibinfo
  {author} {\bibfnamefont {S.}~\bibnamefont {{de Gironcoli}}}, \bibinfo
  {author} {\bibfnamefont {S.}~\bibnamefont {Fabris}}, \bibinfo {author}
  {\bibfnamefont {G.}~\bibnamefont {Fratesi}}, \bibinfo {author} {\bibfnamefont
  {R.}~\bibnamefont {Gebauer}}, \bibinfo {author} {\bibfnamefont
  {U.}~\bibnamefont {Gerstmann}}, \bibinfo {author} {\bibfnamefont
  {C.}~\bibnamefont {Gougoussis}}, \bibinfo {author} {\bibfnamefont
  {A.}~\bibnamefont {Kokalj}}, \bibinfo {author} {\bibfnamefont
  {M.}~\bibnamefont {Lazzeri}}, \bibinfo {author} {\bibfnamefont
  {L.}~\bibnamefont {Martin-Samos}}, \bibinfo {author} {\bibfnamefont
  {N.}~\bibnamefont {Marzari}}, \bibinfo {author} {\bibfnamefont
  {F.}~\bibnamefont {Mauri}}, \bibinfo {author} {\bibfnamefont
  {R.}~\bibnamefont {Mazzarello}}, \bibinfo {author} {\bibfnamefont
  {S.}~\bibnamefont {Paolini}}, \bibinfo {author} {\bibfnamefont
  {A.}~\bibnamefont {Pasquarello}}, \bibinfo {author} {\bibfnamefont
  {L.}~\bibnamefont {Paulatto}}, \bibinfo {author} {\bibfnamefont
  {C.}~\bibnamefont {Sbraccia}}, \bibinfo {author} {\bibfnamefont
  {S.}~\bibnamefont {Scandolo}}, \bibinfo {author} {\bibfnamefont
  {G.}~\bibnamefont {Sclauzero}}, \bibinfo {author} {\bibfnamefont {A.~P.}\
  \bibnamefont {Seitsonen}}, \bibinfo {author} {\bibfnamefont {A.}~\bibnamefont
  {Smogunov}}, \bibinfo {author} {\bibfnamefont {P.}~\bibnamefont {Umari}}, \
  and\ \bibinfo {author} {\bibfnamefont {R.~M.}\ \bibnamefont {Wentzcovitch}},\
  }\bibfield  {title} {\enquote {\bibinfo {title} {Quantum {ESPRESSO}: a
  modular and open-source software project for quantum simulations of
  materials},}\ }\href {\doibase 10.1088/0953-8984/21/39/395502} {\bibfield
  {journal} {\bibinfo  {journal} {J. Phys. Condens. Matt.}\ }\textbf {\bibinfo
  {volume} {21}},\ \bibinfo {pages} {395502} (\bibinfo {year}
  {2009})}\BibitemShut {NoStop}%
\bibitem [{\citenamefont {Chen}\ \emph {et~al.}(2007)\citenamefont {Chen},
  \citenamefont {Fan}, \citenamefont {Tan}, \citenamefont {Wuand},
  \citenamefont {Shu}, \citenamefont {Lu},\ and\ \citenamefont
  {Wang}}]{chen07}%
  \BibitemOpen
  \bibfield  {author} {\bibinfo {author} {\bibfnamefont {N.}~\bibnamefont
  {Chen}}, \bibinfo {author} {\bibfnamefont {L.~Z.}\ \bibnamefont {Fan}},
  \bibinfo {author} {\bibfnamefont {K.}~\bibnamefont {Tan}}, \bibinfo {author}
  {\bibfnamefont {Y.~Q.}\ \bibnamefont {Wuand}}, \bibinfo {author}
  {\bibfnamefont {C.~Y.}\ \bibnamefont {Shu}}, \bibinfo {author} {\bibfnamefont
  {X.}~\bibnamefont {Lu}}, \ and\ \bibinfo {author} {\bibfnamefont {C.~R.}\
  \bibnamefont {Wang}},\ }\bibfield  {title} {\enquote {\bibinfo {title}
  {Comparative spectroscopic and reactivity studies of
  {S}c$_{3-x}${Y}$_x${N}@{C}$_{80}$ ($x = {0-3}$)},}\ }\href {\doibase
  10.1021/jp073229a} {\bibfield  {journal} {\bibinfo  {journal} {J. Phys. Chem.
  C}\ }\textbf {\bibinfo {volume} {111}},\ \bibinfo {pages} {11823--11828}
  (\bibinfo {year} {2007})}\BibitemShut {NoStop}%
\bibitem [{\citenamefont {Dubrovin}\ \emph {et~al.}(2019)\citenamefont
  {Dubrovin}, \citenamefont {Gan}, \citenamefont {B\"{u}chner}, \citenamefont
  {Popov},\ and\ \citenamefont {Avdoshenko}}]{dub19}%
  \BibitemOpen
  \bibfield  {author} {\bibinfo {author} {\bibfnamefont {V.}~\bibnamefont
  {Dubrovin}}, \bibinfo {author} {\bibfnamefont {L-H.}\ \bibnamefont {Gan}},
  \bibinfo {author} {\bibfnamefont {B.}~\bibnamefont {B\"{u}chner}}, \bibinfo
  {author} {\bibfnamefont {A.~A.}\ \bibnamefont {Popov}}, \ and\ \bibinfo
  {author} {\bibfnamefont {S.~M.}\ \bibnamefont {Avdoshenko}},\ }\bibfield
  {title} {\enquote {\bibinfo {title} {Endohedral metal-nitride cluster
  ordering in metallofullerene-{N}i$^\text{II}$({OEP}) complexes and crystals:
  a theoretical study},}\ }\href {\doibase 10.1039/C9CP00634F} {\bibfield
  {journal} {\bibinfo  {journal} {Phys. Chem. Chem. Phys.}\ }\textbf {\bibinfo
  {volume} {21}},\ \bibinfo {pages} {8197--8200} (\bibinfo {year}
  {2019})}\BibitemShut {NoStop}%
\bibitem [{\citenamefont {Yang}\ \emph {et~al.}(2008)\citenamefont {Yang},
  \citenamefont {Popov},\ and\ \citenamefont {Dunsch}}]{yan08}%
  \BibitemOpen
  \bibfield  {author} {\bibinfo {author} {\bibfnamefont {S.~F.}\ \bibnamefont
  {Yang}}, \bibinfo {author} {\bibfnamefont {A.~A.}\ \bibnamefont {Popov}}, \
  and\ \bibinfo {author} {\bibfnamefont {L.}~\bibnamefont {Dunsch}},\
  }\bibfield  {title} {\enquote {\bibinfo {title} {Carbon pyramidalization in
  fullerene cages induced by the endohedral cluster: Non-scandium mixed metal
  nitride clusterfullerenes},}\ }\href {\doibase
  https://doi.org/10.1002/anie.200802009} {\bibfield  {journal} {\bibinfo
  {journal} {Angew. Chem. Int. Ed.}\ }\textbf {\bibinfo {volume} {47}},\
  \bibinfo {pages} {8196--8200} (\bibinfo {year} {2008})}\BibitemShut {NoStop}%
\bibitem [{\citenamefont {Westerstr\"{o}m}\ \emph {et~al.}(2021)\citenamefont
  {Westerstr\"{o}m}, \citenamefont {Dubrovin}, \citenamefont {Junghans},
  \citenamefont {Schlesier}, \citenamefont {B\"uchner}, \citenamefont
  {Avdoshenko}, \citenamefont {Popov}, \citenamefont {Kostanyan}, \citenamefont
  {Dreiser},\ and\ \citenamefont {Greber}}]{Westerstroem21}%
  \BibitemOpen
  \bibfield  {author} {\bibinfo {author} {\bibfnamefont {R.}~\bibnamefont
  {Westerstr\"{o}m}}, \bibinfo {author} {\bibfnamefont {V.}~\bibnamefont
  {Dubrovin}}, \bibinfo {author} {\bibfnamefont {K.}~\bibnamefont {Junghans}},
  \bibinfo {author} {\bibfnamefont {C.}~\bibnamefont {Schlesier}}, \bibinfo
  {author} {\bibfnamefont {B.}~\bibnamefont {B\"uchner}}, \bibinfo {author}
  {\bibfnamefont {S.~M.}\ \bibnamefont {Avdoshenko}}, \bibinfo {author}
  {\bibfnamefont {A.}~\bibnamefont {Popov}}, \bibinfo {author} {\bibfnamefont
  {A.}~\bibnamefont {Kostanyan}}, \bibinfo {author} {\bibfnamefont
  {J.}~\bibnamefont {Dreiser}}, \ and\ \bibinfo {author} {\bibfnamefont
  {T.}~\bibnamefont {Greber}},\ }\bibfield  {title} {\enquote {\bibinfo {title}
  {Precise measurement of angles between two magnetic moments and their
  configurational stability in single-molecule magnets},}\ }\href {\doibase
  10.1103/PhysRevB.104.224401} {\bibfield  {journal} {\bibinfo  {journal}
  {Phys. Rev. B}\ }\textbf {\bibinfo {volume} {104}},\ \bibinfo {pages}
  {224401} (\bibinfo {year} {2021})}\BibitemShut {NoStop}%
\bibitem [{\citenamefont {Nagashima}\ \emph {et~al.}(1995)\citenamefont
  {Nagashima}, \citenamefont {Tejima}, \citenamefont {Gamou}, \citenamefont
  {Kawai},\ and\ \citenamefont {Oshima}}]{nag95}%
  \BibitemOpen
  \bibfield  {author} {\bibinfo {author} {\bibfnamefont {A.}~\bibnamefont
  {Nagashima}}, \bibinfo {author} {\bibfnamefont {N.}~\bibnamefont {Tejima}},
  \bibinfo {author} {\bibfnamefont {Y.}~\bibnamefont {Gamou}}, \bibinfo
  {author} {\bibfnamefont {T.}~\bibnamefont {Kawai}}, \ and\ \bibinfo {author}
  {\bibfnamefont {C.}~\bibnamefont {Oshima}},\ }\bibfield  {title} {\enquote
  {\bibinfo {title} {Electronic structure of monolayer hexagonal boron nitride
  physisorbed on metal surfaces},}\ }\href {\doibase
  10.1103/PhysRevLett.75.3918} {\bibfield  {journal} {\bibinfo  {journal}
  {Phys. Rev. Lett.}\ }\textbf {\bibinfo {volume} {75}},\ \bibinfo {pages}
  {3918--3921} (\bibinfo {year} {1995})}\BibitemShut {NoStop}%
\bibitem [{\citenamefont {Lof}\ \emph {et~al.}(1992)\citenamefont {Lof},
  \citenamefont {van Veenendaal}, \citenamefont {Koopmans}, \citenamefont
  {Jonkman},\ and\ \citenamefont {Sawatzky}}]{lof92}%
  \BibitemOpen
  \bibfield  {author} {\bibinfo {author} {\bibfnamefont {R.~W.}\ \bibnamefont
  {Lof}}, \bibinfo {author} {\bibfnamefont {M.~A.}\ \bibnamefont {van
  Veenendaal}}, \bibinfo {author} {\bibfnamefont {B.}~\bibnamefont {Koopmans}},
  \bibinfo {author} {\bibfnamefont {H.~T.}\ \bibnamefont {Jonkman}}, \ and\
  \bibinfo {author} {\bibfnamefont {G.~A.}\ \bibnamefont {Sawatzky}},\
  }\bibfield  {title} {\enquote {\bibinfo {title} {Band gap, excitons, and
  coulomb interaction in solid {${\mathrm{C}}_{60}$}},}\ }\href {\doibase
  10.1103/PhysRevLett.68.3924} {\bibfield  {journal} {\bibinfo  {journal}
  {Phys. Rev. Lett.}\ }\textbf {\bibinfo {volume} {68}},\ \bibinfo {pages}
  {3924--3927} (\bibinfo {year} {1992})}\BibitemShut {NoStop}%
\bibitem [{\citenamefont {Osterwalder}\ \emph {et~al.}(1996)\citenamefont
  {Osterwalder}, \citenamefont {Greber}, \citenamefont {Aebi}, \citenamefont
  {Fasel},\ and\ \citenamefont {Schlapbach}}]{ost96}%
  \BibitemOpen
  \bibfield  {author} {\bibinfo {author} {\bibfnamefont {J.}~\bibnamefont
  {Osterwalder}}, \bibinfo {author} {\bibfnamefont {T.}~\bibnamefont {Greber}},
  \bibinfo {author} {\bibfnamefont {P.}~\bibnamefont {Aebi}}, \bibinfo {author}
  {\bibfnamefont {R.}~\bibnamefont {Fasel}}, \ and\ \bibinfo {author}
  {\bibfnamefont {L.}~\bibnamefont {Schlapbach}},\ }\bibfield  {title}
  {\enquote {\bibinfo {title} {Final-state scattering in angle-resolved
  ultraviolet photoemission from copper},}\ }\href {\doibase
  10.1103/PhysRevB.53.10209} {\bibfield  {journal} {\bibinfo  {journal} {Phys.
  Rev. B}\ }\textbf {\bibinfo {volume} {53}},\ \bibinfo {pages} {10209--10216}
  (\bibinfo {year} {1996})}\BibitemShut {NoStop}%
\bibitem [{\citenamefont {Puschnig}\ \emph {et~al.}(2009)\citenamefont
  {Puschnig}, \citenamefont {Berkebile}, \citenamefont {Fleming}, \citenamefont
  {Koller}, \citenamefont {Emtsev}, \citenamefont {Seyller}, \citenamefont
  {Riley}, \citenamefont {Ambrosch-Draxl}, \citenamefont {Netzer},\ and\
  \citenamefont {Ramsey}}]{pus09}%
  \BibitemOpen
  \bibfield  {author} {\bibinfo {author} {\bibfnamefont {P.}~\bibnamefont
  {Puschnig}}, \bibinfo {author} {\bibfnamefont {S.}~\bibnamefont {Berkebile}},
  \bibinfo {author} {\bibfnamefont {A.~J.}\ \bibnamefont {Fleming}}, \bibinfo
  {author} {\bibfnamefont {G.}~\bibnamefont {Koller}}, \bibinfo {author}
  {\bibfnamefont {K.}~\bibnamefont {Emtsev}}, \bibinfo {author} {\bibfnamefont
  {T.}~\bibnamefont {Seyller}}, \bibinfo {author} {\bibfnamefont {J.~D.}\
  \bibnamefont {Riley}}, \bibinfo {author} {\bibfnamefont {C.}~\bibnamefont
  {Ambrosch-Draxl}}, \bibinfo {author} {\bibfnamefont {F.~P.}\ \bibnamefont
  {Netzer}}, \ and\ \bibinfo {author} {\bibfnamefont {M.~G.}\ \bibnamefont
  {Ramsey}},\ }\bibfield  {title} {\enquote {\bibinfo {title} {Reconstruction
  of molecular orbital densities from photoemission data},}\ }\href {\doibase
  10.1126/science.1176105} {\bibfield  {journal} {\bibinfo  {journal}
  {Science}\ }\textbf {\bibinfo {volume} {326}},\ \bibinfo {pages} {702--706}
  (\bibinfo {year} {2009})}\BibitemShut {NoStop}%
\bibitem [{\citenamefont {Greif}\ \emph {et~al.}(2013)\citenamefont {Greif},
  \citenamefont {Castiglioni}, \citenamefont {Seitsonen}, \citenamefont {Roth},
  \citenamefont {Osterwalder},\ and\ \citenamefont {Hengsberger}}]{gre13}%
  \BibitemOpen
  \bibfield  {author} {\bibinfo {author} {\bibfnamefont {M.}~\bibnamefont
  {Greif}}, \bibinfo {author} {\bibfnamefont {L.}~\bibnamefont {Castiglioni}},
  \bibinfo {author} {\bibfnamefont {A.~P.}\ \bibnamefont {Seitsonen}}, \bibinfo
  {author} {\bibfnamefont {S.}~\bibnamefont {Roth}}, \bibinfo {author}
  {\bibfnamefont {J.}~\bibnamefont {Osterwalder}}, \ and\ \bibinfo {author}
  {\bibfnamefont {M.}~\bibnamefont {Hengsberger}},\ }\bibfield  {title}
  {\enquote {\bibinfo {title} {Photoelectron diffraction in the x-ray and
  ultraviolet regime: {S}n-phthalocyanine on {A}g(111)},}\ }\href {\doibase
  10.1103/PhysRevB.87.085429} {\bibfield  {journal} {\bibinfo  {journal} {Phys.
  Rev. B}\ }\textbf {\bibinfo {volume} {87}},\ \bibinfo {pages} {085429}
  (\bibinfo {year} {2013})}\BibitemShut {NoStop}%
\end{thebibliography}
\end{document}